\documentclass[review]{elsarticle}
\usepackage{lineno,hyperref}
\usepackage{floatrow}
\usepackage{caption}
\usepackage{graphicx}
\usepackage{multicol,multirow}
\usepackage{amsmath,amssymb,amsfonts}
\usepackage{mathrsfs}
\usepackage{amsthm}
\usepackage{subfig}
\usepackage[hmargin=2cm]{geometry}

\modulolinenumbers[5]

\usepackage{xcolor}

\begin{document}

\begin{frontmatter}
\title{Incomplete to complete multiphysics forecasting - a hybrid approach for learning unknown phenomena} 

\author[1]{Nilam Tathawadekar\corref{correspondingauthor}}
\cortext[correspondingauthor]{Corresponding author}
\ead{nilam.tathawadekar@tum.de}
\author[2]{Nguyen Anh Khoa Doan}
\author[3]{Camilo F. Silva}
\author[1]{Nils Thuerey}

\address[1]{Department of Informatics, Technical University of Munich, Boltzmannstr. 3, 85748 Garching, Germany}
\address[2]{Faculty of Aerospace Engineering, Delft University of Technology, Kluyverweg 1, 2629HS Delft, Netherlands}
\address[3]{Department of Mechanical Engineering, Technical University of Munich, Boltzmannstr. 15, 85747 Garching, Germany}

\begin{abstract}
Modeling complex dynamical systems with only partial knowledge of their physical mechanisms is a crucial problem across all scientific and engineering disciplines. Purely data-driven approaches, which only make use of an artificial neural network and data, often fail to accurately simulate the evolution of the system dynamics over a sufficiently long time and in a physically consistent manner. Therefore, we propose a hybrid approach that uses a neural network model in combination with an incomplete partial differential equations (PDE) solver that provides known, but incomplete physical information. In this study, we demonstrate that the results obtained from the incomplete PDEs can be efficiently corrected at every time step by the proposed hybrid neural network – PDE solver model, so that the effect of the unknown physics present in the system is correctly accounted for. For validation purposes, the obtained simulations of the hybrid model are successfully compared against results coming from the complete set of PDEs describing the full physics of the considered system. We demonstrate the validity of the proposed approach on a reactive flow, an archetypal multi-physics system that combines fluid mechanics and chemistry, the latter being the physics considered unknown. Experiments are made on planar and Bunsen-type flames at various operating conditions. The hybrid neural network - PDE approach correctly models the flame evolution of the cases under study for significantly long time windows, yields improved generalization, and allows for larger simulation time steps.
\end{abstract}


\end{frontmatter}

\section{Introduction}\label{sec:intro}
Modeling and forecasting of complex physical systems described by nonlinear partial differential equations (PDEs) are central to various domains with applications ranging from weather forecasting \citep{kalnay2003atmospheric}, design of airplane wings \citep{rhie1983numerical,zafar2021recurrent}, to material science \citep{wheeler1992phase}. Typically, a chosen set of PDEs are solved iteratively until convergence of the solution. 
Modeling complex physical dynamics requires a good understanding of the underlying physical phenomena. For cases where the complete information on the physics of the system is missing, deep learning models can be employed to complete the physical description when additional data of the system is available. Deep learning methods have shown promise to account for these unknown components of the system \citep{um2020sol,yin2021augmenting}. 
We consider a set of partial differential equations with partially unknown physics represented.
The corresponding PDE model for a general state $\phi$ is given by

\begin{equation}\label{eq:generic}
\begin{aligned}
    \frac{\partial \phi}{\partial t} &= \mathcal{P}_c\left(\phi, \frac{ \partial \phi}{\partial x}, \frac{\partial^2 \phi}{\partial x^2}, ...\right), & \quad \text{in}\quad \Omega \times (0,\infty) \\
    &= \mathcal{F}\left(\mathcal{P}_{i}\left(\phi, \frac{ \partial \phi}{\partial x}, \frac{\partial^2 \phi}{\partial x^2}, ...\right), \mathcal{P}_u \left(\phi, \frac{ \partial \phi}{\partial x}, \frac{\partial^2 \phi}{\partial x^2}, ...\right)\right), & \quad \text{in}\quad \Omega \times (0,\infty) \\
    \mathcal{G}(\phi) &= 0, & \quad \text{in}\quad \partial\Omega \times (0,\infty) \\
    \phi &= h, & \quad \text{in}\quad \Omega \times (0) \\
\end{aligned}
\end{equation}
where $\mathcal{P}_c$ represents the physical system with complete information. $\mathcal{F}$ represents a potentially simple function that combines the known but incomplete PDE description, $\mathcal{P}_i$, and unknown physics represented by $\mathcal{P}_u$ to match the solutions of complete PDE system. We take $\mathcal{G}$ and $h$ to be known functions appropriately defining the boundary and initial conditions respectively. $\Omega$ is the spatial domain over which we solve the PDE system and $\partial \Omega$ its boundary. The term $\mathcal{P}_u$ can take the form of closure terms, source terms, higher order coupling terms between state variables or terms resulting from a set of unknown ODEs/PDEs depending on the physical system under investigation.
\begin{figure}[bt!]
\centering
\includegraphics[width=0.9\linewidth]{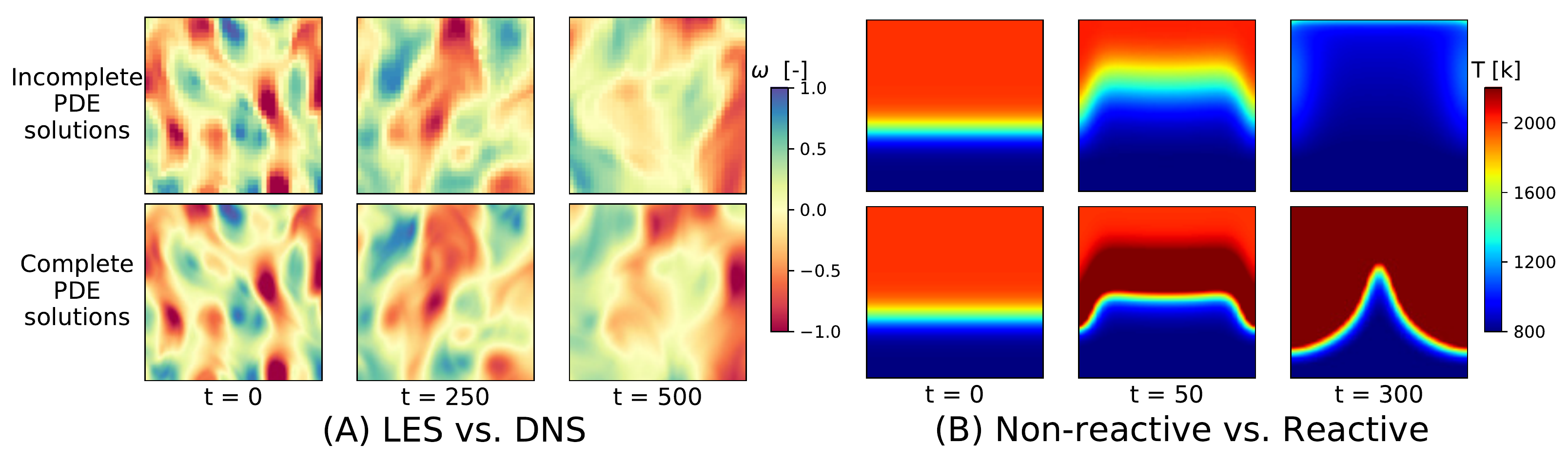}
\caption{(A) The normalized vorticity solutions of complete/DNS (bottom) solver can be reached by increasing spatial resolution of the incomplete/LES (top) solver \citep{list2022learned}. (B) We consider the problem of the incomplete/non-reactive (top) and complete/reactive (bottom) PDE solvers which can yield fundamentally different evolutions, as shown here for a sample temperature field over time}
\label{fig:reactive_flow}
\end{figure}

A commonly targeted case is when the governing equations of the complete PDE description are computationally too expensive to solve, with turbulence modeling in computational fluid dynamics (CFD) being a good example \citep{pope2000turbulent,chen1997fundamentals}.
In CFD, a spatial filtering is performed on the original governing PDEs in the context of large eddy simulations (LES). This step introduces unclosed terms in the model equations that correspond to unrepresented physics in equation \eqref{eq:generic}, due to the effects of the filtered scales. Figure \ref{fig:reactive_flow} (A) shows instances of normalized vorticity for isotropic decaying turbulence. The solutions of a fully resolved direct numerical simulation (DNS) could be achieved by increasing the spatial resolution of LES. 
This is a widely studied problem, where the use of deep learning models is currently being explored \citep{lapeyre2019training, stachenfeld2022learned, kochkov2021machine, list2022learned}. 
Several of these methods train a neural network with LES as the incomplete system to model the effects of the filtered scales and obtain the solutions of the complete DNS  \citep{kochkov2021machine,list2022learned} . 
Instead, we consider a different, more challenging problem related to multi-physics, coupled systems containing dependent variables which describe different physical phenomena. In equation \eqref{eq:generic}, $\mathcal{P}_u$ contains all the terms corresponding to different physical phenomena that are not included in the PDEs represented by $\mathcal{P}_i$.
This formulation could describe fluid-structure interactions which couple fluid mechanics with structural mechanics with $\mathcal{P}_u$ representing the unknown coupling terms, or aeroacoustics problems which couple fluid dynamics and acoustics \citep{benra2011comparison,liu2018multiphysics}. 

However, in this work, we will focus on reactive flow simulation as the complete PDE description, while a non-reactive flow simulation will represent the incomplete PDE basis. It collects the chemical kinetic mechanisms in the unknown physics term of equation \eqref{eq:generic}. In this system, $\mathcal{P}_u$ could take the form of the source terms from unknown chemistry in the Navier-Stokes equations. A \textit{reactive flow} refers to a fluid flow with chemical reactions occurring within a reacting fluid, such as combustion-related flows. A reacting fluid is a mixture of two or more species such as hydrocarbons, oxygen, water, carbon dioxide, etc which undergo chemical reactions \citep{poinsot2005theoretical}. In contrast, a non-reactive flow refers to a fluid flow where no chemical reactions take place. Figure \ref{fig:reactive_flow} (B) shows a visual example of a non-reactive and reactive flow simulations which can be considered as an incomplete and complete PDE systems, respectively. Starting from the same initial condition, the influence of $\mathcal{P}_u$ in this multi-physics system leads to fundamentally different solutions as reacting fluid (shown by blue color in figure \ref{fig:reactive_flow}(B)) advances through the domain without reacting for the non-reactive flow simulation or forms a $\Lambda$ shaped flame with higher temperature of burned products (shown by dark red color) for the reactive flow simulation. 
Therefore, we are targeting a more challenging problem than those tackled in \citep{lapeyre2019training, kochkov2021machine, list2022learned, stachenfeld2022learned}, where increasing spatial resolution and/or reducing time-scales of the incomplete PDE solver does not lead to a converged full solution. 
Rather, the complete and incomplete PDEs produce drastically different solutions due to the unknown physics. The central learning objective is to correct this behavior and retrieve the evolution that would be obtained with the complete PDE description.

Our work expands on the combination of incomplete PDE solvers and neural networks (NNs)
\citep{yin2021augmenting,takeishi2021physics}
to account for the effects of an incomplete physics model.
The NN aims to complete the PDE description, where the differences in complete and incomplete PDE solutions are beyond the effects of spatial and temporal scales. 
We showcase that combining the trained NN model with a differentiable solver for the incomplete PDE can accurately reproduce the physical solutions of the complete, multi-physics PDE solver with stable long-term rollouts.

Reactive flow modeling has applications in numerous domains such as combustion processes in gas turbines \citep{lieuwen2012unsteady,gruber2018direct}, climate modeling \citep{jacobson1999fundamentals,rolnick2022tackling} and astrophysics simulations \citep{gamezo2003thermonuclear}.
Resolving the Navier-Stokes equations lies at the core of these problems, where additionally the transport of different species of relevance must also be accounted for, together with their production or consumption often following complex reaction mechanisms \citep{poinsot2005theoretical}. For chemically reacting flows, generation or consumption of multiple species via some chemical reaction are modeled using a net source term. It is a well-known fact that the incorporation of a detailed chemical kinetic mechanism in a reacting flow model can result in a stiff system of governing equations \citep{wanner1996solving, najm1998semi, knio1999semi}. 

We showcase the effectiveness of our approach for different cases of planar 2D premixed methane-air flames, and the varying transient evolution of Bunsen-type flames. We show that the proposed approach can handle large domains with highly resolved flames, which are closer to the practical flame domains used in many industrial applications. Specifically we concentrate on training a NN model to correct the spatio-temporal effects of energy and species transport source terms. We show that in addition to recovering the desired solutions, this approach overcomes inherent problems of temporal stiffness due to the complex reaction mechanism. Lastly, we show that the resulting  hybrid solver provides a flexible building block for adjacent tasks. Specifically, we demonstrate this for controlling the evolution of flame shapes via continuous control. The hybrid solvers can efficiently predict the states of the reactive flows and readily enable the control of the flow inlet velocities to arrive at a desired flame shape. Therefore, we employ a pre-trained hybrid solver to train a second controller network that learns to steer the flame simulation such that the observed states are matched.

This paper is organized as follows. We discuss the relevant literature in Section \ref{related_work}. Section \ref{sec:method} describes the problem statement, the mathematical formulation, details of the differentiable PDE solver and neural network model. Discussion on data generation process is provided in Section \ref{sec:data}. Section \ref{sec:experiments} shows the results obtained on different planar and Bunsen flame scenarios considered. Section \ref{sec:discussion} demonstrates the robustness of the proposed approach. Finally in Section \ref{sec:flame_shape}, we apply the proposed hybrid approach to arrive at the desired flame shape by solving an optimization problem. Section \ref{sec:conclusion} concludes this paper with a summary of the results and outlines future work.

\section{Related work}
\label{related_work}
Deep learning methods have been widely used to model the solutions of partial differential equations \citep{lagaris1998artificial,long2018pde,han2018solving,bar2019learning} and in particular, the Navier-Stokes equations \citep{guo2016convolutional,bhatnagar2019prediction}.

\textbf{Purely data-driven models:} \quad A simple approach in modeling any physical system consists of training a deep learning model using data coming from either experiments or numerical simulation \citep{thuerey2020deep,fukami2019super}. These models solely use deep learning techniques with appropriate data to make predictions and hence are called as \emph{purely data-driven models}. \cite{thuerey2020deep} applied a convolutional neural network (CNN) to learn Reynolds-Averaged Navier-Stokes (RANS) solutions of airfoil flows. The proposed approach is  very generic and applicable to a wide range of PDE boundary value problems on Cartesian grids. \cite{stachenfeld2022learned} demonstrates that a generic CNN-based models may predict turbulent dynamics on coarse grids more accurately than classical numerical solvers. The proposed approach is effectively applied on a wide range of physical domains which can be represented as grids. These classical neural networks map between finite-dimensional spaces and can only learn solutions tied to a specific discretization which can be excessively limiting. Therefore, approaches based on learning operators are receiving increased attention.  \cite{lu2021learning} proposed a novel architecture based on fully connected neural networks called DeepONets to learn diverse linear or nonlinear explicit and implicit operators.
Neural operators \citep{li2020neural, bhattacharya2021model, patel2021physics}, specifically, the Fourier neural operator (FNO) of \cite{li2020fourier} introduce an interesting line of work by learning mesh-free, infinite-dimensional operators with neural networks, but do not necessarily offer advantages for longer term predictions. We compare the proposed approach with such data-driven models as baselines.
Purely data-driven models can be very fast and may not suffer from the time-step stability issues associated with traditional numerical solvers. Nevertheless, as these purely data-driven approaches lack the physical understanding of the system being modeled, they generally fail in generalizing to other operating conditions \citep{kim2019deep,lapeyre2019training}. To leverage the potential of deep learning in physical simulations, it is therefore necessary to incorporate some physical information within the deep learning framework.  

\textbf{Physics-informed deep learning models:} \quad Deep learning models can enforce physical constraints partially through the loss function \citep{bar2019learning,raissi2019physics,yadav2022physics} or changes in neural network architecture \citep{greydanus2019hamiltonian}. 
A well-known framework called Physics-Informed Neural Network (PINN) \citep{raissi2019physics} uses neural networks as methods for solving PDEs. It minimizes the residual of the underlying governing laws by taking advantage of automatic differentiation to compute exact, mesh-free derivatives. However, these approaches struggle to enforce physical constraints such as boundary conditions or predict the strong unsteadiness and chaotic nature of flows \citep{dwivedi2020physics,fuks2020limitations}. PINN only learns the solution function of a single PDE instance and needs re-optimization for other instances or PDEs. To alleviate this issue, \cite{wang2021learning} extended the DeepONet framework by imposing the underlying physical laws via soft penalty constraints during training. Although physics-informed DeepONet imposes PDE losses on operator learning, they are not discretization invariant. To tackle this issue, physics-informed neural operator (PINO) framework was proposed by \cite{li2021physics} that uses available data and physics constraints to learn the solution operator of parametric PDEs. However, all these approaches require the explicit knowledge of the underlying PDEs to accurately train the network. For the physical systems described in equation \eqref{eq:generic} where only partial knowledge of their physical mechanism is known, these approaches may fail to converge to the solutions of the complete PDE solver. Minimizing the residual of only the known but incomplete PDEs will not lead to an accurate prediction of the solutions of a complete PDE system. Additionally, these approaches would need some further modifications to be implemented for the problem considered. Therefore, they are not considered as a baseline method for the present work.  

\textbf{Neural networks with differentiable PDE solvers:} \quad In recent years, the development of differentiable PDE solvers have led to an interesting line of research. \cite{thuerey2021physics} have developed an open-source physics simulation toolkit called \emph{PhiFlow} for optimization and machine learning applications. \emph{SU2} \citep{economon2016su2} is an open-source collection of software tools for analysing PDEs and PDE-constrained optimization problems on unstructured meshes with state-of-the-art numerical methods. \emph{SciML} \citep{rackauckas2020universal} is a collection of tools for solving equations and modeling systems. These tools provide differentiable functions for physical simulations, which enable close integration with deep learning frameworks by leveraging their automatic differentiation functionality. Hybrid approaches that combine machine learning techniques with numerical PDE solvers \citep{wang2020towards, illarramendi2022performance}, have attracted a significant amount of interest due to their capabilities for generalization \citep{chen2018neural}. In this context, neural networks are typically used to model or replace a part of the conventional PDE solver to improve aspects of the solving process. For example, \cite{tompson2017accelerating}, \cite{ozbay2019poisson} and \cite{ajuria2020towards} proposed a convolutional neural network based approach to solve the Poisson equation in CFD simulation. In recent years, a number of deep learning based models have been introduced to accurately model turbulent flows \citep{pathak2020using, stachenfeld2022learned, dresdner2022learning}. \cite{um2020sol} and \cite{belbute2020combining} showed the advantages of training neural networks with differentiable physics to correct the numerical errors that arise in the discretization of PDEs. \cite{sirignano2020dpm} introduced a deep learning PDE augmentation method. Large eddy simulation of turbulence is augmented with a deep neural network to model unresolved physics to obtain direct numerical simulation solutions. 
\cite{kochkov2021machine} correct for closure error by integrating a neural network with a differentiable CFD simulator. These approaches demonstrate the capabilities of neural networks to correct errors in a fast, under-resolved simulation. The unresolved physics in turbulence modeling is attributed to spatial filtering. Given additional compute resources, one can arrive at accurate solutions by running the known PDEs at higher resolution. Note that in this work we investigate a more challenging case of unknown physics in multi-physics system, where increasing the spatial or temporal resolution of the known but incomplete PDEs will not lead to the accurate solution of the complete PDEs.

Similar to the goals of our work,
\cite{yin2021augmenting} and \cite{takeishi2021physics} have introduced frameworks of augmenting incomplete physical dynamics with neural network models. 
These approaches demonstrate the applicability on ODE/PDE systems, which are weakly nonlinear and the unknown dynamics are linear combinations of the underlying flow fields. 
We expand on these works to explore the significantly more challenging scenario of reactive flows. These are characterised by a multi-physics system with nonlinear advective terms which models the transport of a flow state by the velocity of the flow, and strongly nonlinear dynamics, described by exponential source terms that exhibit nonlinear combinations of the flow fields. Compared to the work by \cite{yin2021augmenting}, we do not use an additive approach to learn the solutions of the complete PDE system. Instead, we use the output of an incomplete PDE solver as an input to the neural network model to correct for the effects of unknown terms.

\section{Methodology}\label{sec:method}
\begin{figure}[bt!]
\centering
\includegraphics[width=0.9\linewidth]{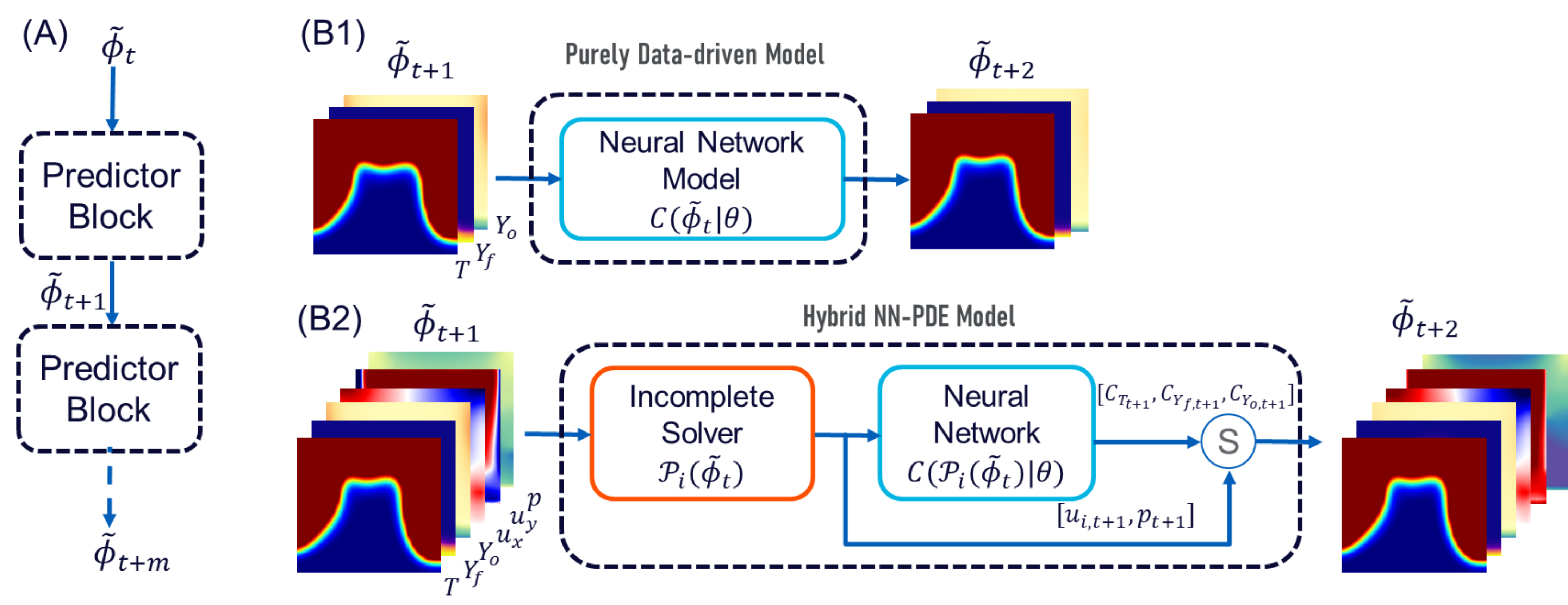}
\caption{(A) Multi-step training framework helps to learn the dynamics of complete PDE solver over longer rollouts. (B1) Details of the input flow state and predictor block used in a purely data-driven approach and (B2) the hybrid NN-PDE approach, where S denotes the concatenation of different fields to obtain the complete flow state $\tilde{\phi}$ at next time step}
\label{fig:solvernn}
\end{figure}
We consider two different sets of PDEs with their associated numerical solver, which we denote as the {\em incomplete} PDE $\mathcal{P}_{i}$ and the {\em complete} PDE $\mathcal{P}_{c}$. By evaluating $\mathcal{P}_c$ on an input state $\phi_t$ at time $t$, we can compute the points of the phase space sequences; $\phi_{t+\Delta t} = \mathcal{P}_c(\phi_t)$. Without loss of generality, we assume a fixed time-step $\Delta t$ and denote a state $\phi_{t+\Delta t}$ at next time instance as $\phi_{t+1}$. 
Let $\mathcal{X}$ be a Banach space of functions taking values in a spatial domain $\Omega \subset \mathbb{R}^2$. Furthermore, let $C^{\dagger} : \mathcal{X} \rightarrow \mathcal{X} $ be a nonlinear map from $\phi_t$ to $\phi_{t+1}$, defined over a finite time interval $[0, T]$ and an input flow state $\phi$. The learning objective is to find the best possible correction function 
\begin{equation}
    C(\phi;\theta) : \mathcal{X} \rightarrow \mathcal{X}, \quad \theta \in \Theta
\end{equation}
for some finite dimensional parameter space $\Theta$ by choosing $\theta^{\dagger}$ such that $C(.;\theta^{\dagger}) \approx C^{\dagger}$. The neural network models can be employed to learn such mapping as shown in figure \ref{fig:solvernn}.
The model parameters $\theta$ are estimated from the complete PDE solution trajectories $(\phi_0, \phi_1, .., \phi_T)$.
The learned predictions obtained after repeatedly applying the corrector $C$ and invoking $\mathcal{P}_{i}$ are denoted by $(\tilde{\phi_0}, \tilde{\phi_1}, .., \tilde{\phi_T})$. 

\subsection{Partial differential equations for reactive flows}\label{sec:pdes}
In this section, we present the basics of reactive flow simulation and underlying conservation equations. A reactive flow involves multiple species reacting through one or more chemical reactions. Different species, such as hydrocarbons (fuel), oxygen (oxidizer), water, carbon dioxide (products), are characterized through their mass fraction $Y_k$. The primary variables for two-dimensional (2-D) reacting flow involve density ($\rho$), 2-D velocity field ($u$), temperature ($T$), and the mass fraction $Y_k$ of the $N$ reacting species. 

In a premixed combustor, fuel ($Y_F$) and oxidizer ($Y_O$) are mixed before they enter the combustion chamber. The computation of premixed flames with complex chemistry is possible but we consider a simplified approach. We assume that chemistry proceeds only through one irreversible reaction i.e. one-step chemistry. If $\upsilon_F^\prime$ and $\upsilon_O^\prime$ are the coefficients corresponding to fuel and oxidizer when considering a one-step reaction of type $\upsilon_F^\prime F + \upsilon_O^\prime O \rightarrow \text{products},$ the mass fractions of fuel and oxidizer correspond to stoichiometric conditions. It is defined as, 
\begin{equation}
    \left(\frac{Y_O}{Y_F}\right)_{st} = \frac{\upsilon_O^\prime W_O}{\upsilon_F^\prime W_F} = \varphi.
\end{equation}
This ratio $\varphi$ is called the mass stoichiometric ratio. $W_F$ and $W_O$ represent the molecular weights of the fuel and oxidizer, respectively. The equivalence ratio of a given mixture is then, 
\begin{equation}
    E = \varphi \left(\frac{Y_F}{Y_O}\right).
\end{equation}
A common example of such reaction is $CH_4 + 2O_2 \rightarrow CO_2 + 2H_2O$, where $\upsilon_F^\prime = 1, \upsilon_O^\prime = 2, W_F = 0.016 \text{ kg/mole}, W_O=0.032 \text{ kg/mole}$ and therefore $\varphi =4$. The equivalence ratio is an important parameter in the design of a premixed combustion system. Rich combustion is observed for $E > 1$ (the fuel is in excess) and lean regimes are achieved when (E < 1) (the oxidizer is in excess) \citep{poinsot2005theoretical}. Most practical premixed combustors operate at or below stoichiometry \citep{lieuwen2005combustion}.

The physical system we investigate is a laminar premixed methane-air flame with one-step chemistry. It is governed by the following Navier-Stokes equations \citep{poinsot2005theoretical}
\begin{equation}\label{ns_equations}
\begin{aligned}
\frac{\partial \rho}{\partial t} + \nabla . (\rho u) &= 0  &\quad \text{in}\quad \Omega \times [0,T]\\
\frac{\partial}{\partial t} (\rho u) + \nabla . (\rho u \otimes u) &= - \nabla p + \nabla . \tau &\quad \text{in}\quad \Omega \times [0,T] \\ 
\rho C_p \left(\frac{\partial T}{\partial t} + \nabla . (u \otimes T) \right)  &= \dot{\omega}_T^\prime + \lambda {\nabla}^2 T &\quad \text{in}\quad \Omega \times [0,T] \\
\frac{\partial \rho Y_k}{\partial t} + \nabla . (\rho u \otimes Y_k)  &= \dot{\omega}_k + \rho D_k {\nabla}^2 Y_k &\quad \text{in}\quad \Omega \times [0,T] \\
\end{aligned}
\end{equation}
where $\tau$, $D_k$, $\lambda$ are the strain rate tensor, the diffusion coefficient of species $k$, and the mixture thermal conductivity. In addition, $C_p$ denotes the mixture specific heat capacity. $\dot{\omega}_k$ and $\dot{\omega}_T^\prime$ are the species reaction rate and heat release rate, respectively. Boundary conditions of each flow state vary depending on the applications and are provided in detail in section \ref{sec:data}. $\Omega \subset \mathbb{R}^2$ represents the spatial domain.
 
The reaction rate $\dot{\omega}_k$ for each species are linked to the progress rate $Q_1$ at which the single reaction proceeds, as: $\dot{\omega}_k = W_k \upsilon_k Q_1$. The simplifications proposed by \cite{williams1985combustion} and \cite{mitani1980propagation} are used to model the reaction rates $\dot{\omega}_k$ for each species. The progress rate $Q_1$ is assumed to have the Arrhenius form and is given by,
\begin{equation}
\label{eq:omegaF}
\begin{aligned}
    Q_1 &= B_1 T^{\beta_1} \left(\frac{\rho Y_F}{W_F}\right)^{n_F} \left(\frac{\rho Y_O}{W_O}\right)^{n_O} \exp\left(-\frac{E_a}{RT}\right). \\
    \dot{\omega}_F &= \upsilon_F^\prime  W_F Q_1; \quad \dot{\omega}_O = \upsilon_O^\prime  W_O Q_1
\end{aligned}
\end{equation}
where $\dot{\omega}_F$ and $\dot{\omega}_O$ are the reaction rates of fuel and oxidizer, respectively.
$\dot{\omega}_T^\prime $ is the heat release due to combustion and is formulated as, $\dot{\omega}_T^\prime = - \sum_{k=1}^N \Delta h_{f,k}^{o} \dot{\omega}_k$. The formation enthalpy $h_{f,k}^{o}$ is the enthalpy needed to form 1 kg of species $k$ at the reference temperature $T_0 = 298.15 K$. The formulation for $\dot{\omega}_T^\prime$ can be further simplified as,
\begin{equation}\label{eq:omegaT}
    \dot{\omega}_T^\prime = - \sum_{k=1}^N \Delta h_{f,k}^{o} W_k \upsilon_k Q_1 = - \sum_{k=1}^N \Delta h_{f,k}^{o} \frac{W_k \upsilon_k}{W_F \upsilon_F} W_F \upsilon_F  Q_1 = - \sum_{k=1}^N \Delta h_{f,k}^{o} \frac{W_k \upsilon_k}{W_F \upsilon_F} \dot{\omega}_F = - Q \dot{\omega}_F.
\end{equation}

Therefore, the heat release source term $\dot{\omega}_T^\prime$ and the fuel source term $\dot{\omega}_F$ are linked by the $Q$, which is the heat of reaction per unit mass. Following \cite{poinsot2005theoretical}, parameters corresponding to a real-world methane-air flame are chosen as: $ B_1 = 1.08 10^7 \text{ uSI}; \quad \beta_1 = 0; \quad E_a = 83600  \text{ J/mole}; \quad n_F = 1; \quad n_O = 0.5; \quad Q = 50100 \text{ kJ/kg}; \quad C_p = 1450 \text{ J/(kgK)} $.
Taken together, the system of equations above is a challenging scenario even for classical solvers, and due to its practical relevance likewise a highly interesting environment for deep learning methods.

\subsection{Problem formulation}\label{sec:problem}
The incomplete PDE solver solves the set of equations \eqref{ns_equations} without the source terms and reaction rates $\dot{\omega}_k$ and $\dot{\omega}_T^\prime$, while the complete PDE solver solves the full set of equations \eqref{ns_equations}. The neural network model, denoted by $C(\mathcal{P}_{i}(\phi)|\theta)$, corrects the incomplete/non-reacting flow states $\mathcal{P}_{i}(\phi)$ to obtain the complete/reacting states $\mathcal{P}_{c}(\phi)$ as shown in figure \ref{fig:solvernn} (B2). 
The neural network is trained to model the effects of the unknown chemistry using parameters $\theta$ given an input flow state, $\phi =[u_i, p, T, Y_f, Y_o]$.
As seen from equation \eqref{ns_equations}, equation \eqref{eq:omegaF}, and equation \eqref{eq:omegaT}, the temperature and species mass fraction fields are strongly coupled, which significantly increases the prediction problem complexity. A slight error in one of the fields will quickly propagate into the other fields, making the predictions diverge. In the following, a subscript $C_{s}(\circ)$ will denote that the neural network $C$ generates the field $s$, e.g., $C_{T}$ generating the temperature field. The update can be written as, 
$\tilde{\phi}_{t+1} = [u_{i,t+1}, p_{t+1}, C_T(\mathcal{P}_{i}(\tilde{\phi}_t)|\theta), C_{Y_f}(\mathcal{P}_{i}(\tilde{\phi}_t)|\theta), C_{Y_o}(\mathcal{P}_{i}(\tilde{\phi}_t)|\theta)]$
where $\tilde{\cdot}$ indicates a corrected state. $u_{i,t+1}$ and $p_{t+1}$ are the time-advanced velocity and pressure field, respectively, predicted using the incomplete PDE solver. 

\subsection{Training Methodology}\label{sec:models}
\begin{figure}[bt!]
\centering
\includegraphics[width=0.9\linewidth]{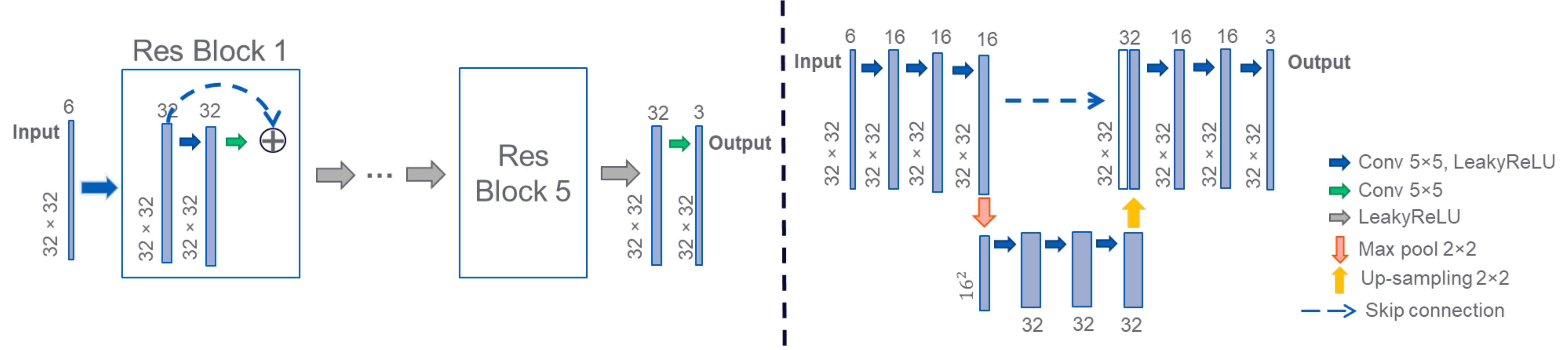}
\caption{Schematic of the convolutional neural networks used. Left: ResNet with 5 ResBlocks, Right: UNet32 with 2 layers}
\label{fig:res_unet}
\end{figure}

\textbf{Hybrid NN-PDE approach} \quad We employ a hybrid NN-PDE approach that augments a neural network model with a PDE solver \citep{um2020sol,kochkov2021machine}.
In contrast to previous work, we use the incomplete PDE solver as a basis, and hence the solver does not converge to the desired solutions under refinement, as explained in section \ref{sec:intro}.
The neural network is integrated and trained in a loop with the incomplete PDE solver using stochastic gradient descent for $m$ time steps, as shown in figure \ref{fig:solvernn} (A,B2). Here, the number of temporal look-ahead steps, $m$, is an important hyper-parameter of the training process. Higher $m$ provides the network with longer-term feedback at training time through the gradient roll-outs. This gives the model improved feedback on how the time dynamics of the incomplete PDE solver affect the input states, and hence which corrections need to be inferred by the model. 
In our framework, if an incomplete PDE solver is very close to the complete PDE solver for the given data, the NN model would learn a correction which is closer to an identity. The skip connections in the used convolutional neural network architectures would readily enable such an identity operation. In contrast, when the incomplete solver contributes very little, the proposed approach would start to represent a purely data-driven approach.

\textbf{Differentiable PDE Solver} \quad A central component of the 
hybrid NN-PDE model is the differentiable solver, which allows us to embed the solver for the incomplete PDE system in the training of a neural network. The differentiable solver acts as additional non-trainable layers in the network. 
They provide derivatives of the outputs of the simulation with respect to its inputs and parameters.
Finite differences can be used to compute the gradients of the PDE solver, but they are  computationally very expensive for high-dimensional PDEs. 
Differentiable solvers resolve this issue by solving the adjoint problem \citep{pontryagin1987mathematical} via analytic derivatives.
Here, we use the differentiable PDE solver from the {\em phiflow} framework of \cite{thuerey2021physics} in combination with {\em Tensorflow} to obtain the non-reactive flow solver and reactive flow solver solutions.
The marker-and-cell method \citep{harlow1965numerical} is adopted to represent temperature, pressure, density, species mass fraction fields in a centered grid, and velocities in a staggered grid. Basic differential operators such as gradient, divergence, curl, and Laplace operators are implemented in TensorFlow using basic mathematical tensor operations \citep{holl2020phiflow}. 
These differential operators act on a 9 point stencil of grid points and the corresponding derivatives are straight-forward to compute. Advection is implemented with a semi-Lagrangian step, while the Poisson problem for the pressure and its gradient are solved implicitly. Efficient derivatives for all these operations are then combined via back-propagation \citep{werbos1990backpropagation}.

\textbf{Purely data-driven (PDD) approach} \quad A PDD model is used as a baseline. 
It employs a neural network model to learn the complete flow states $\mathcal{P}_c(\phi)$ given an input flow state $\phi^S$ where $\phi^S = [T, Y_f, Y_o]$, as shown in figure \ref{fig:solvernn} (B1). The new state predicted by the trained neural network model is $\tilde{\phi}^{S}_{t+1} = C(\tilde{\phi}^{S}_t|\theta)$. For the hybrid NN-PDE as well as PDD, the neural network part of the predictor block in figure \ref{fig:solvernn} consists of a fully convolutional neural network model. 

We additionally compare against the \emph{Fourier neural operator} (FNO) of \cite{li2020fourier} as an example of a state-of-the-art neural operator method. 
The new state predicted by the trained model is given by $\tilde{\phi}^{S}_{t+1} = C(\tilde{\phi}^{S}_t|\theta)$ where $\tilde{\phi}^S = [T, Y_f, Y_o]$ and $C(\circ)$ represents the Fourier neural operator.

\subsection{Training Details}\label{sec:training_details}
\begin{figure}
	\centering
	\includegraphics[width=0.5\linewidth]{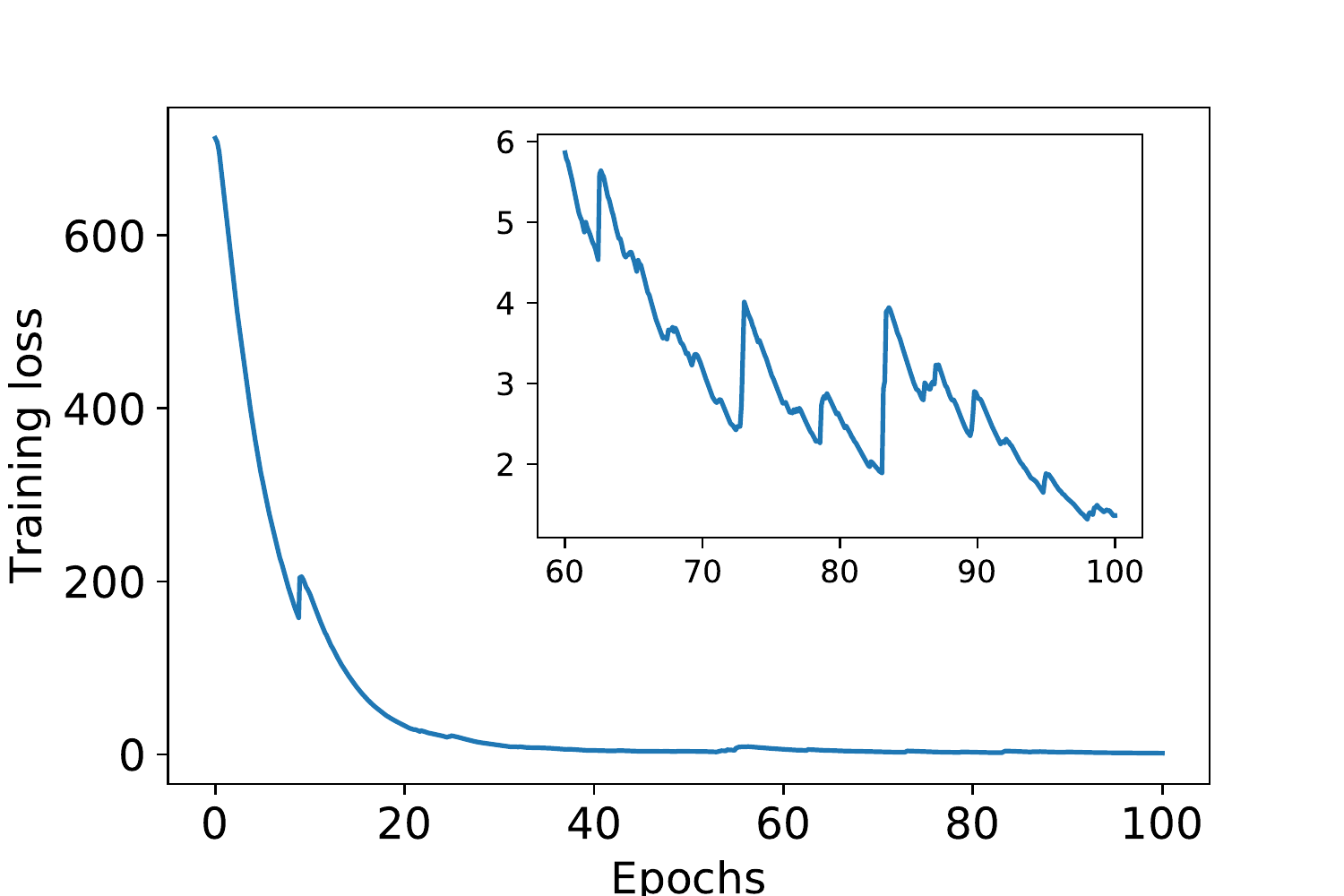}
	\caption{Typical $L_2$ based training loss as defined in equation \eqref{eq:loss} over 100 epochs. The inset figure shows zoomed in loss function curve for last 40 epochs}
	\label{fig:loss}
\end{figure}

All three approaches use an $L_2$ based loss that is evaluated for $m$ steps as
\begin{equation}\label{eq:loss}
\mathcal{L}(\theta) = \sum_{n=t+1}^{t+m} \sum_{\phi = \{T, Y_f, Y_o\}} || \tilde{\phi}_n, \mathcal{P}_{c}(\phi_{n}) ||_2 \ .
\end{equation}
The network receives the input states as shown in figure \ref{fig:solvernn}. The output of the neural network model is used to obtain the corrected state $\tilde{\phi}_{t+1}$ as specified above. We constrain the mass fraction fields $Y_k$ to contain physical values in the range $Y_f \in [0,0.05]$ and $Y_o \in [0,1]$. All models are trained for 100 epochs with a batch size of 3 and a learning rate of 0.0001. We have used a small batch size of 3 due to the memory requirements as we unroll the NN-PDE framework over long timesteps. Our training procedure uses the Adam optimizer \citep{kingma2015adam}. For all our computations, a \emph{Nvidia Quadro RTX 8000} GPU is used. Figure \ref{fig:loss} shows the typical training loss curve over 100 epochs.

For PDD and the hybrid NN-PDE approach, we experimented with both the ResNet \citep{he2016deep} and the UNet \citep{ronneberger2015u} architectures. Figure \ref{fig:res_unet} shows the schematic of the neural network models used. We found that the ResNet performed the best for the PDD setting, while for the hybrid NN-PDE approach the UNet performed consistently better. Hence, the following results will use a ResNet for PDD models, and a UNet for the NN-PDE hybrids. 
Additional details of the neural network architectures are provided in section \ref{app:NN} of the supplementary material as well as results comparing the UNet and ResNet architectures.

\section{Numerical Experiments}\label{sec:data}
\begin{figure}[bt!]
\centering
\includegraphics[width=0.65\linewidth]{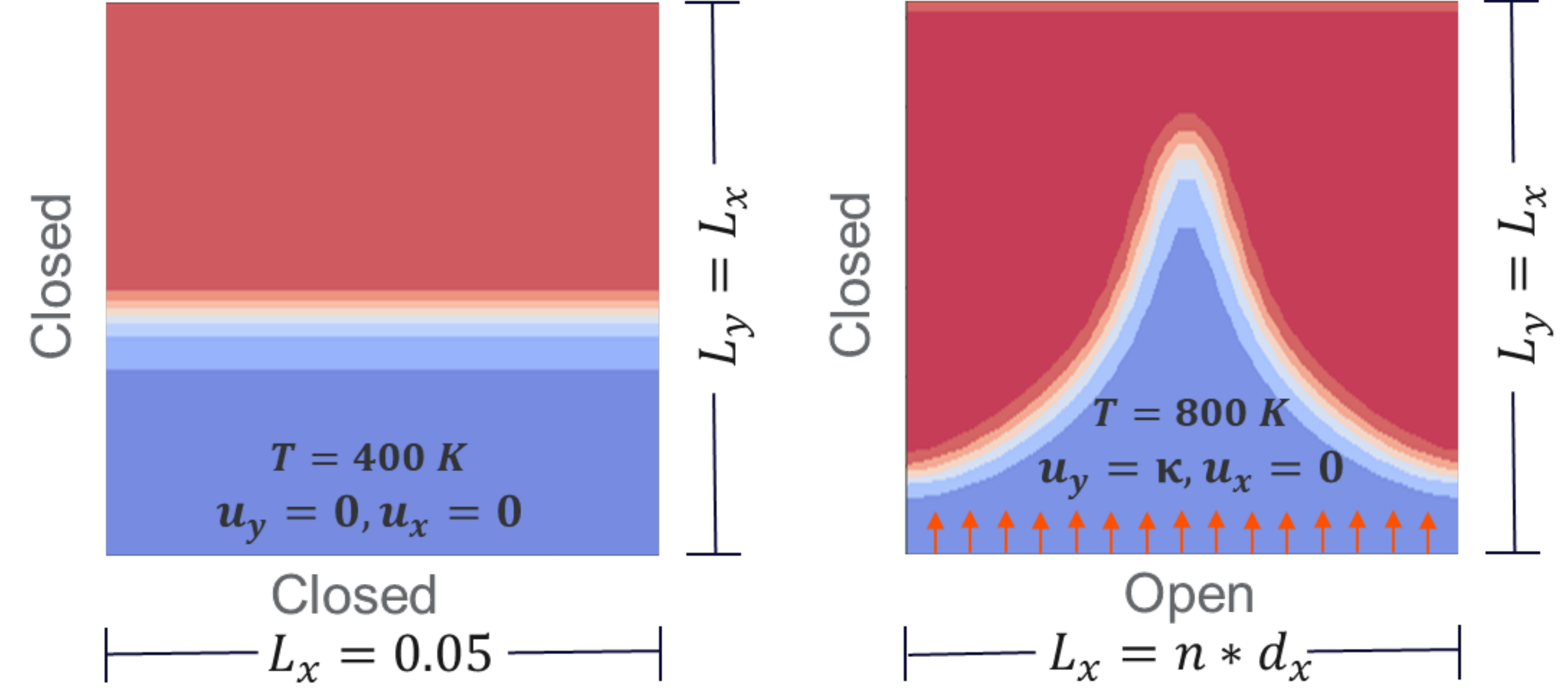}
\caption{Details of the boundary condition for - left: Planar flame; right: a Bunsen-type flame. n represents resolution of the domain}
\label{fig:bc}
\end{figure}

\begin{table}[bt!]
\caption{Details of the boundary conditions used for the planar flame case and various cases of Bunsen-type flames. The uniform-Bunsen case is obtained with $\kappa$ = constant and nonUniform-Bunsen case is obtained with $\kappa  \in \mathbb{R}^n$ as discussed in section \ref{sec:data}.}
\centering
\begin{tabular}{ c c }   
Planar flame & Bunsen-type flame \\  
\begin{tabular}{ c c c } 
\hline
field & inlet & left and right wall\\  
\hline
 T & 400 K & Neumann BC \\ 
 $u_y$ & 0 & - \\
 $u_x$ & 0 & - \\
 $p$ & 101325 & Neumann BC \\
 $Y_f$ & $\frac{1}{1 + \frac{\varphi}{E} (1 + 3.76 \frac{W_{N_2}}{W_{O_2}})}$ & slip \\
\hline
\end{tabular} &  
\begin{tabular}{ c c c } 
\hline
field & inlet & left and right wall\\  
\hline
 T & 800 K & Neumann BC \\ 
 $u_y$ & $\kappa$ & no-slip \\
 $u_x$ & 0 & slip \\
 $p$ & 101325 & Neumann BC \\
 $Y_f$ & $\frac{1}{1 + \frac{\varphi}{E} (1 + 3.76 \frac{W_{N_2}}{W_{O_2}})}$ & slip \\
\hline
\end{tabular} \\
\end{tabular}
\label{table:bc}
\end{table}

\begin{figure}[bt!]
\centering
\includegraphics[width=0.99\linewidth]{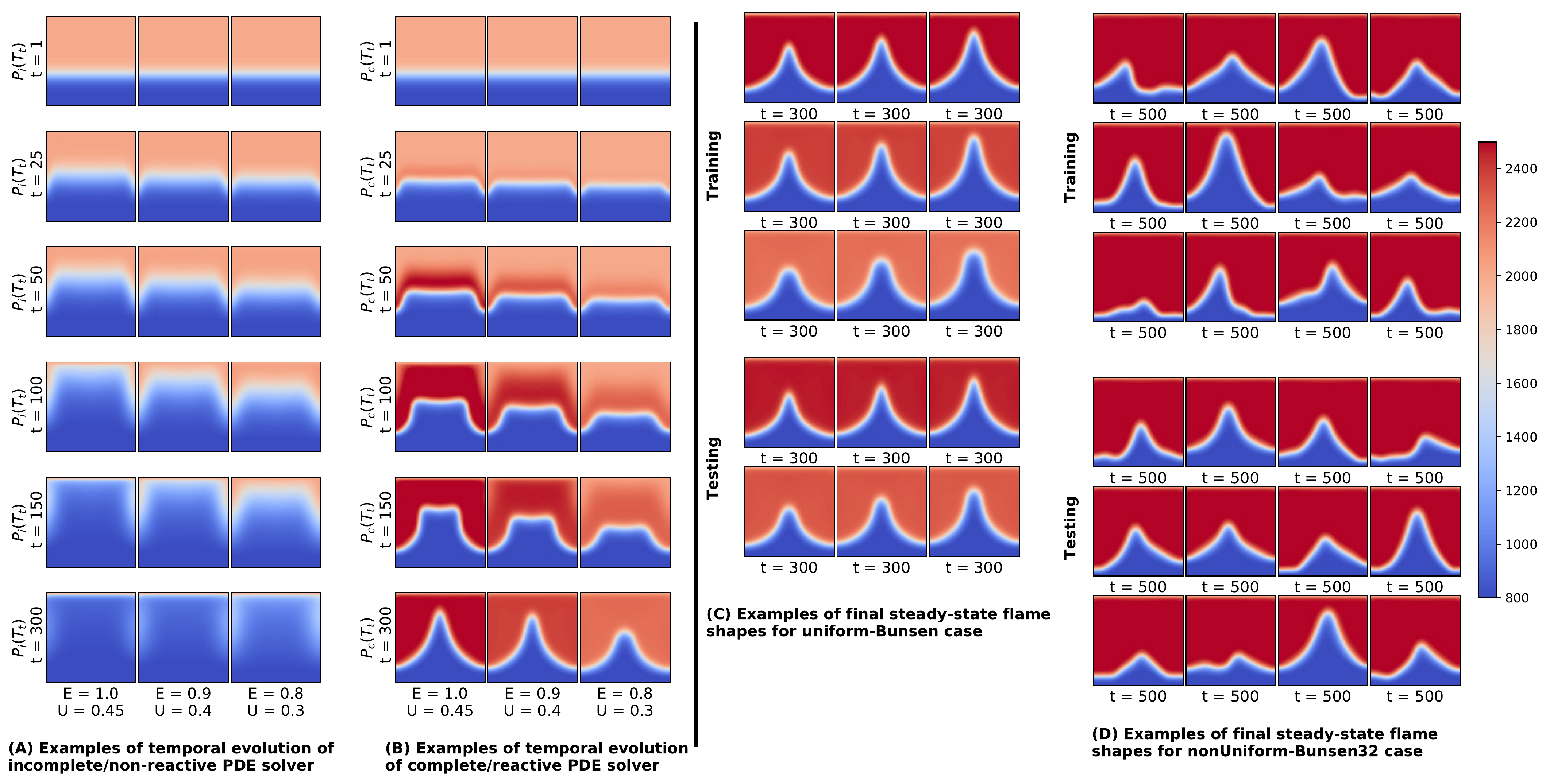}
\caption{Instances of temporal evolution of (A): incomplete PDE solver; (B): complete PDE solver for different operating conditions. (C): Snapshot of complete PDE solver at $t = 300$ for all training and testing datasets of the uniform-Bunsen case. (D): Snapshot of complete PDE solver at $t = 500$ for all training and testing datasets of the nonUniform-Bunsen32 case}
\label{fig:all_train_test}
\end{figure}

We consider a case of planar 2D premixed methane-air flame propagating in a quiescent mixture (\textbf{Planar-v0}) and two cases of the transient evolution of an initially planar laminar premixed flame into a Bunsen-type flame under different inlet velocity conditions (\textbf{uniform-Bunsen, nonUniform-Bunsen}). 
We obtain the target data by considering the source terms as defined by equation \eqref{eq:omegaF} and equation \eqref{eq:omegaT} with the parameters mentioned in section \ref{sec:pdes}.

\textbf{Planar-v0 \quad} For the most basic case, the planar 2D flame model setup, we consider the reacting Navier-Stokes equations described in section \ref{sec:pdes} with zero inlet velocity i.e. $u_x = 0$, $u_y = 0$. We consider a square domain of size 0.05 m $\times$ 0.05 m with 32 $\times$ 32 resolution and closed boundary conditions, as shown in figure \ref{fig:bc}. 
The simulation is initialized using a steep transition between a premixed methane-air mixture and burnt gases. Our training data consists of 6 simulations of 300 steps created by varying the equivalence ratio
$E$. It represents the stoichiometric mixture ($\varphi$) of fuel $Y_f$ and oxidizer $Y_o$ mass fractions, i.e., $E = \varphi \frac{Y_f}{Y_o}$ and thus fundamentally influences the dynamics of the chemical reaction.
For the training data we use
$E_{\text{train}} = \{1.0, 0.9, 0.8, 0.7, 0.6 \}$, while the test dataset contains $E_{\text{test}} = \{0.95, 0.85, 0.75, 0.65\}$. 

\textbf{uniform-Bunsen \quad} In contrast to the planar case, the premixed methane-air mixture is now fed with a constant inlet velocity. The boundary conditions upstream, at $y = 0$, are $(u_x,u_y)_{x,y=0} = (0,\kappa)$ where $\kappa$ is a $n$-dimensional vector with constant amplitude. The target simulation contains a heat release rate term, and in this case, the initial temperature field evolves into different $\Lambda$-shaped flames at the end of the 300$^{th}$ time step. Training and testing datasets are created by varying the equivalence ratio and inlet velocity amplitude $U$: 
$E_{\text{train}} = \{1.0, 0.9, 0.8\}$ and $U_{\text{train}} = \{0.45, 0.4, 0.3\}$. The test dataset uses $E_{\text{test}} = \{0.95, 0.85\}$ $U_{\text{test}} = \{0.43, 0.375, 0.325\}$. The length and temperature of the flame significantly vary depending on the inlet velocity and equivalence ratio provided. 
Figure \ref{fig:all_train_test} (A) shows the temperature field evolution of the incomplete PDE solver at different time instances for 3 different operating conditions for the uniform-Bunsen case. Figure \ref{fig:all_train_test} (B) shows the corresponding target training data for the uniform-Bunsen case. All the simulations are run for 300 steps. It shows the difference between non-reactive and reactive flow solver simulations for different operating conditions over 300 time-steps. Figure \ref{fig:all_train_test} (C) shows the last snapshot ($t = 300$) of 9 training simulations and 6 testing datasets with operating parameters interpolated between the training data parameters for uniform-Bunsen case. These datasets are obtained by varying the equivalence ratio and the magnitude of the uniform inlet velocity.
The flame temperature depends on the equivalence ratio used and the flame height depends on the inlet velocity amplitude and equivalence ratio.

\textbf{nonUniform-Bunsen \quad} As a third case we consider the transient evolution of a premixed methane-air flame with non-uniform inlet velocity. The boundary conditions upstream, at y = 0, are $(u_x,u_y)_{x,y=0} = (0,\kappa)$ where $\kappa$ is a $n$-dimensional vector whose elements are each sampled from a uniform distribution from [0.2, 0.65]. We experiment with two different domain sizes, 32 $\times$ 32 (\textbf{nonUniform-Bunsen32}) and 100 $\times$ 100 (\textbf{nonUniform-Bunsen100}) .
The larger domain size used is closer to the practical reactive flow domain utilized in CFD applications \citep{jaensch2017hybrid} with a highly resolved flame. These inlet velocity conditions generate complex flame shapes, which increase the difficulty of the prediction problem. We consider simulation sequences with 500 steps, as it takes longer time for the flame to reach the steady-state solution. Figure \ref{fig:all_train_test} (D) showcases the snapshots of training and testing dataset at $t = 500$ for nonUniform-Bunsen32 case. Non-uniform variations in inlet velocity profile leads to different complex flame shapes. We use 12 datasets with 500 simulation steps to train the models and test it on 12 test cases shown in figure \ref{fig:all_train_test} (D). For the nonUniform-Bunsen32 case with 32 unrolling steps ($m=32$), training requires approximately 60 hours with approximately 1 GB of GPU memory.

To generate input and target data for training, we simulate the temperature $T$, mass fractions $Y_f$, $Y_o$ and velocity $u_x, u_y$ fields of all flames under study. Table \ref{table:bc} summarizes the boundary conditions applied for the planar and Bunsen-type flame cases discussed above. No-slip boundary conditions are used for the velocity field $u_y$ to obtain the $\Lambda$-shaped flames.

The training and test datasets are split using different initial conditions. 
This ensures that varied (training) and unseen (test) data is provided. For example, in the case of Planar-v0 scenario, the training data consists of 5 simulations of planar flame with different equivalence ratios, evolving over 300 simulation steps. During test time, only the initial state of the flow field ($\phi_0$) and unseen boundary conditions (i.e. other equivalence ratios than those in the training datasets) are specified and the trained hybrid NN-PDE model is used to make predictions over 300 time steps. A similar setup is used for the baseline approaches. Multi-step training framework shown in figure \ref{fig:solvernn} demonstrates the training setup for the baseline and hybrid NN-PDE approaches. Continuous time-slices are used during training. As shown in figure \ref{fig:solvernn} (A), during training, the \emph{Predictor Block} is unrolled for $m$ steps i.e. the network is trained using multiple “sequences” of $m$ consecutive snapshots (selected from the cases in the training datasets). During testing, only $\phi_0$ (the initial flow state) along with unseen boundary conditions are provided and the trained predictor block is evaluated 300 times to obtain predictions from $t=1,2…,300$.

\section{Results}
\label{sec:experiments}
We demonstrate the capabilities of the proposed learning approach to represent the complete PDE description with the aforementioned cases of increasing difficulty.
We also study its ability to generalize to unseen operating conditions such as equivalence ratios, simultaneous variations in constant inlet velocity and equivalence ratio, and non-uniform inlet velocity profiles. As baselines, we compare against a purely data-driven approach; a neural network model with exactly the same look-ahead steps, and the Fourier neural operator of Li et al. (2020a) that likewise includes $m$ look-ahead steps as discussed in section \ref{sec:models}. All the qualitative and quantitative evaluations shown in the paper are performed on test datasets with unseen initial conditions.

\subsection{Planar-v0}
\begin{figure}[bt!]
\centering
\includegraphics[width=0.8\linewidth]{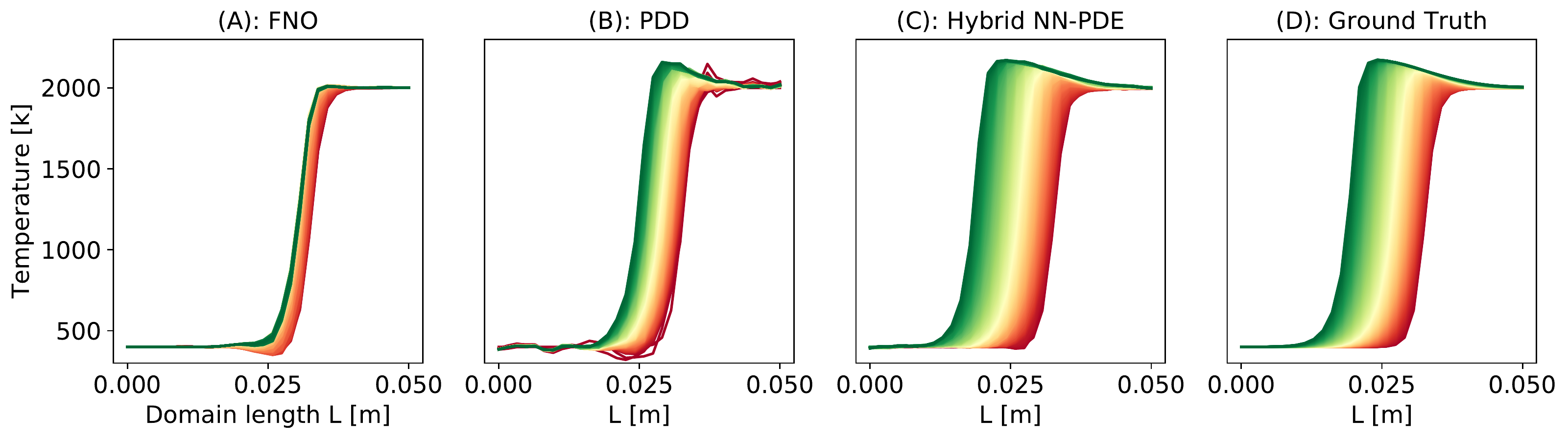}
\caption{1D cut of the planar flame simulation over 300 steps. The initial state is plotted in red, target state in green. Hybrid NN-PDE approach predicts physically accurate results over longer rollouts.}
\label{fig:CR_fig1}
\end{figure}

\begin{figure}[bt!]
\centering
\includegraphics[width=0.8\linewidth]{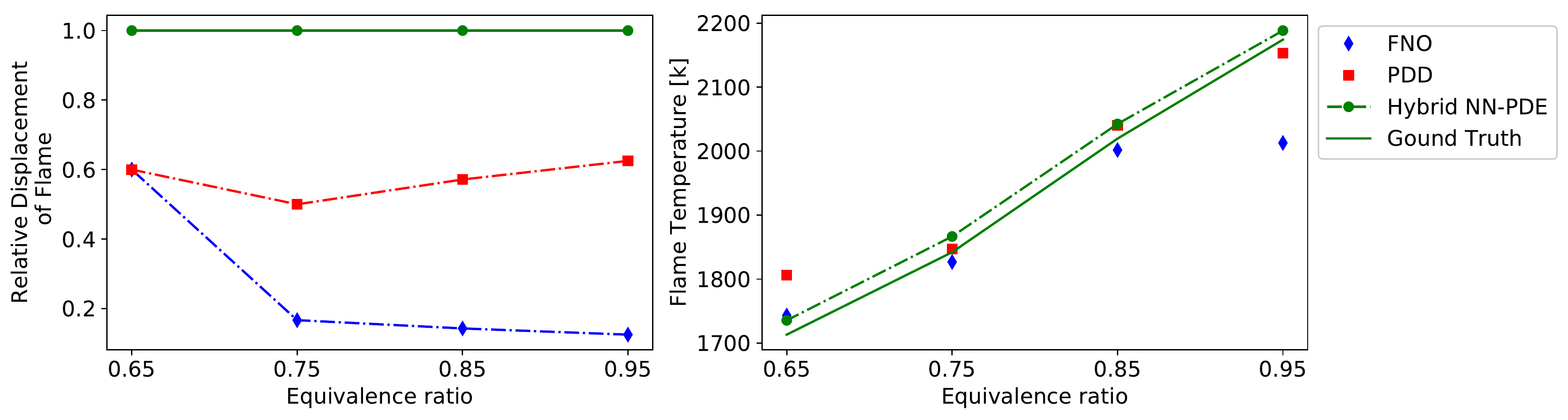}
\caption{Hybrid NN-PDE approach predicts physically accurate results with correct flame temperature and relative displacement of flame front across different equivalence ratios}
\label{fig:CR_params}
\end{figure}

\begin{table}[bt!]
\caption{Mean and standard deviation of errors over all time steps of all testsets. The Hybrid NN-PDE approach outperforms all other baselines considered.}
\label{tab:mape_mse}
\begin{center}
\resizebox{\textwidth}{!}{
\begin{tabular}{ c|c|c|c|c|c| } 
\cline{2-6}
\\[-1em]
& & Baseline & Baseline & Hybrid & Hybrid \\
& & FNO & NN & NN-PDE & NN-PDE-dt \\
\cline{2-6}
\\[-1em]
\parbox[t]{2mm}{\multirow{4}{*}{\rotatebox[origin=c]{90}{MAPE}}} & Planar-v0 & 8.27 $\pm$ 5.51\% & 6.33 $\pm$ 3.05\% & 1.40 $\pm$ 0.65\% & \textbf{1.21} $\pm$ 0.39\% \\
& uniform-Bunsen & 15.57 $\pm$ 8.72\% & 7.58 $\pm$ 3.73\% & \textbf{0.72} $\pm$ 0.37\% & 1.11 $\pm$ 0.47\% \\
& nonUniform-Bunsen32 & 12.30 $\pm$ 7.98\% & 12.48 $\pm$ 11.31\% &  \textbf{2.04} $\pm$ 1.39\% & 2.46 $\pm$ 1.61\%   \\
& nonUniform-Bunsen100 & - & - & \textbf{3.23} $\pm$ 3.76\% & 4.14 $\pm$ 4.99\% \\
\cline{2-6}
\\[-1em]
\parbox[t]{2mm}{\multirow{4}{*}{\rotatebox[origin=c]{90}{MSE}}} & Planar-v0 & 88316 $\pm$ 124394 & 34491 $\pm$ 43966 & 1122 $\pm$ 1300 & \textbf{292} $\pm$ 296 \\
& uniform-Bunsen & 220817 $\pm$ 167024 & 72563 $\pm$ 56919 & \textbf{721} $\pm$ 1007 & 1267 $\pm$ 1622 \\
& nonUniform-Bunsen32 & 184860 $\pm$ 185694 & 130219 $\pm$ 142094 & \textbf{9647} $\pm$ 13903 & 17569 $\pm$ 24392  \\
& nonUniform-Bunsen100 & - & - & \textbf{23293} $\pm$ 37451 & 26862 $\pm$ 47639 \\
\cline{2-6}
\end{tabular}}
\end{center}
\end{table}

Table \ref{tab:mape_mse} compares the mean absolute percentage errors (MAPE) and mean squared errors (MSE) of temperature field for all the cases discussed in section \ref{sec:data}. For Planar-v0, the FNO and PDD approaches yield large errors with a MAPE of 8.27$\%$ and 6.33$\%$, respectively. 
On the other hand, the hybrid NN-PDE model trained with 32 look-ahead steps reduces the error to 1.4$\%$, and thus performs significantly better than the two baselines. 
This behavior is visualized in figure \ref{fig:CR_fig1} with a 1D transverse cut of the simulation domain over 300 time-steps. The hybrid NN-PDE approach successfully captures the propagation of the flame. 

In figure \ref{fig:CR_params}, we also compare two important physical quantities: the flame temperature and relative displacement of the flame front, across different equivalence ratios. It can be seen that the hybrid NN-PDE model (green circles) accurately predicts the flame temperature for different test cases (solid green line). The relative displacement of the flame front is computed as $|\Tilde{x}_{t} - \Tilde{x}_0|/|x_{t} - x_0|$, where $x_t$ is the position along the flame normal at time $t$ on the $1200 K$ isotherm of the ground truth simulation, and $\Tilde{x}$ denotes the predicted position. For all test cases, the hybrid approach accurately predicts the flame displacement over $t$ = 300 steps, while the other approaches yield significant errors. 
\begin{figure}
\centering
\includegraphics[width=0.9\linewidth]{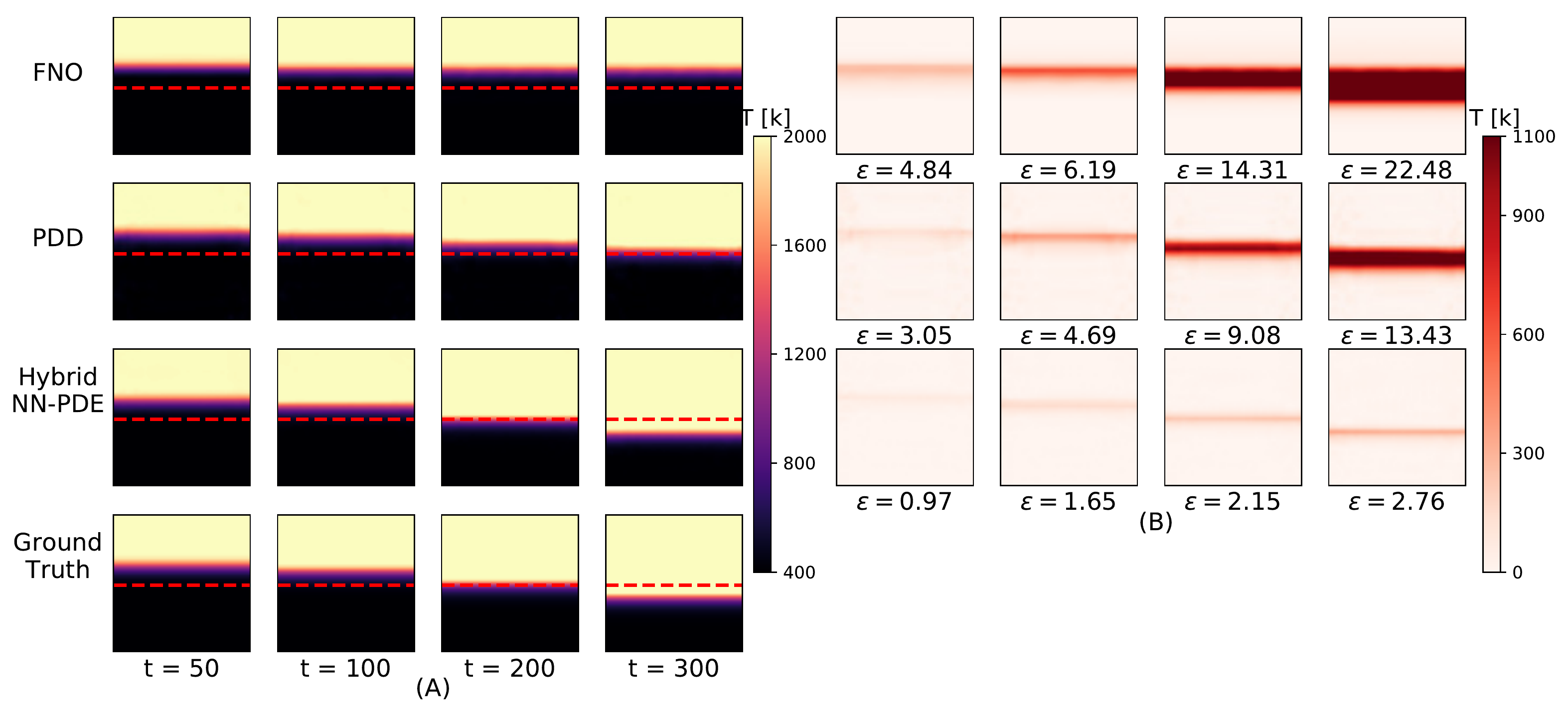}
\caption{Planar-v0 flame case with $E$ = 0.95. (A) Temperature field predictions; (B) Absolute error between ground truth ($\mathcal{P}_{c}(T_{t})$) and the output predicted - from top to bottom top - by: FNO, purely data-driven approach and hybrid NN-PDE. The numbers represent instantaneous MAPE}
\label{fig:cr_combined}
\end{figure}
Figure \ref{fig:cr_combined}(A) shows the 2D visualization of the temperature field predictions for Planar-v0 case. As seen from the ground truth images, methane-air mixture (black color) converts into the burned products (yellow color) due the chemical reaction at the flat flame surface (red color). The dotted, horizontal red line helps to compare the transition of the flame interface, i.e. the displacement of the flame front. Due to the chemical reaction, fuel-air mixture is consumed and turns into burned product as the simulation progresses. The FNO approach completely fails to predict the propagation of the flame for the given test case. Its output does not show any evolution from the initial temperature profile for the given operating condition. The purely data-driven approach fails to capture the flame front displacement correctly, thus leading to an inaccurate prediction with large errors. The hybrid NN-PDE model accurately captures this evolution of planar methane-air flame in a quiescent mixture. Figure \ref{fig:cr_combined}(B) shows the instantaneous MAPE w.r.t. ground truth data for predictions shown in figure \ref{fig:cr_combined}(A). 
The absolute error shown in figure \ref{fig:cr_combined}(B) exceeds 1100 K for the FNO and purely data-driven approaches as these do not predict the flame temperature and flame front displacement correctly. We use the upper limit of 1100 K for colorbar to highlight the errors in the hybrid NN-PDE approach more clearly. Large errors in FNO and PDD results stem from their inability to reliably predict the flame temperature and flame front displacement.

\subsection{Bunsen-type flame}
\begin{figure}[bt!]
\centering
\includegraphics[width=0.9\linewidth]{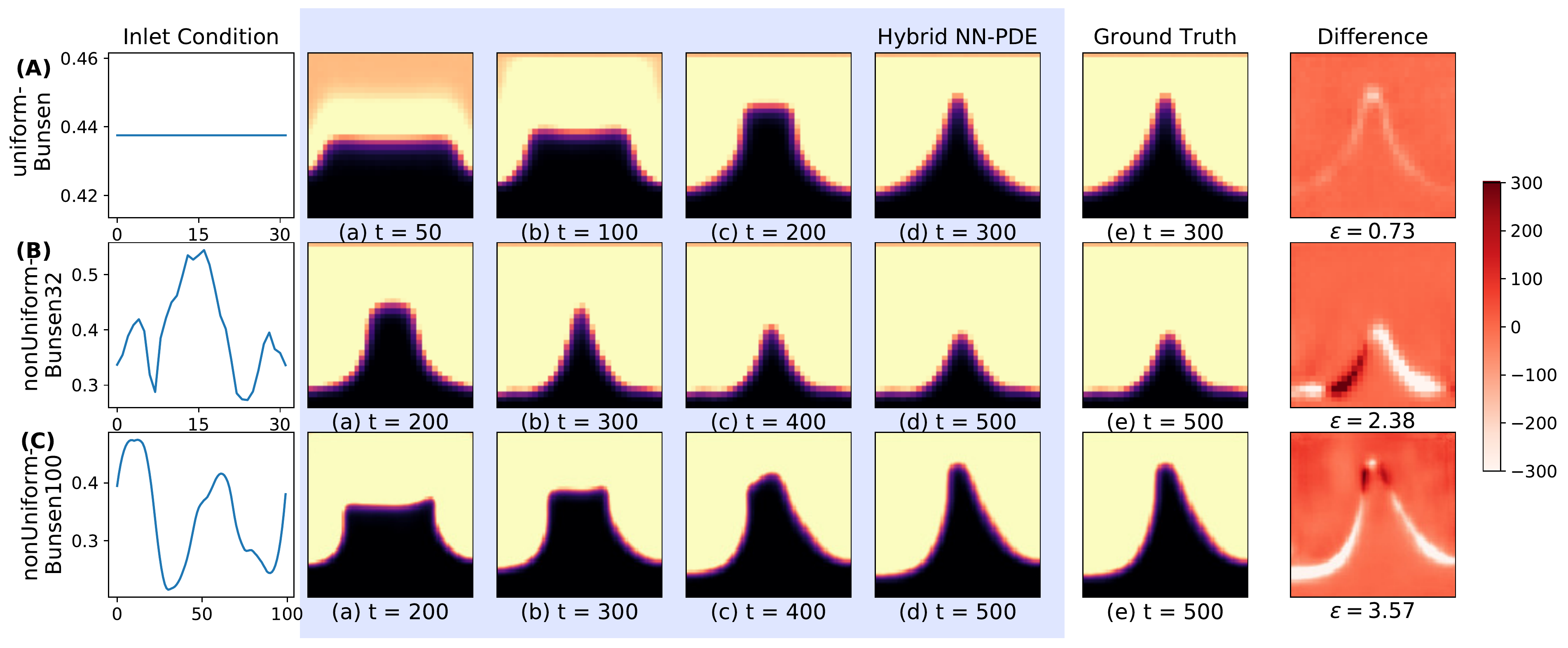}
\caption{Left : Inlet velocity profile used. (a-d) Temperature field prediction by hybrid NN-PDE approach over different steps given the inlet velocity profile. (e) Ground truth data. Right : Difference between ground truth data and hybrid NN-PDE output at last snapshot. 
Top to bottom: 32 $\times$ 32 resolution cases of (A) uniform-Bunsen, (B) nonUniform-Bunsen32, and (C) 100 $\times$ 100 testcase nonUniform-Bunsen100. $\epsilon$ represents the instantaneous MAPE}
\label{fig:FL_32_fig1}
\end{figure}
For the uniform-Bunsen case, the PDD baseline with an error of 7.58$\%$ performs better than the FNO model which yields an error of 15.57$\%$. However, a neural network model combined with an incomplete PDE solver for 32 look-ahead steps yields a significantly lower error of 0.72$\%$  as shown in table \ref{tab:mape_mse}. This means that the hybrid approach reduces the errors by a factor of 10 over the baselines considered. Figure \ref{fig:FL_32_fig1}(A) shows the temporal evolution of the hybrid NN-PDE approach predicted over 300 time steps for the uniform-Bunsen case. It shows the results for a test case with $U = 0.4375$ and $E=0.95$. It predicts a symmetric flame with accurate flame height and achieves very low instantaneous MAPE of 0.73$\%$ at $t$=300, when compared with the ground truth data.

Next, we study a complex scenario of nonUniform-Bunsen flame with 32 $\times$ 32 resolution. The hybrid NN-PDE approach outperforms the PDD approach and FNO with an improvement of $\sim80\%$.
We also include variant of the PDD approach (PDD-5) which is trained to predict all quantities ($\phi^S = [T, Y_f, Y_o, u, p]$) of the flow state (details in section \ref{app:PDD_5} of the supplementary material).
The hybrid NN-PDE model achieves a low MAPE of 2.04\% compared to the higher errors of 23.45\% and 12.48\% predicted by the PDD-5 and the PDD approach, respectively.
The MSE values show this trend even more clearly. The large standard deviations of the MSE numbers indicate that the predictions made by the baseline approaches contain substantial deviations from the target values. 
It is important to highlight that despite using the same training data and a similar neural network architecture with same number of look-ahead steps, the hybrid approach outperforms the PDD approach due to its learned collaboration with the incomplete PDE solver. It accurately reproduces the complete PDE behavior. Figure \ref{fig:FL_32_fig1}(B) highlights this with a visualization of the hybrid NN-PDE model predictions for nonUniform-Bunsen32 case. The trained model accurately predicts the flame simulation over long roll-outs of 500 steps and achieves the complex flame shape with a low, instantaneous MAPE of 2.38$\%$. 

Finally, we showcase the ability of the hybrid approach to predict the temporal evolution of highly resolved flames with the uniform-Bunsen100 scenario. Despite the increased complexity of the larger resolution, it achieves a very good overall MAPE of 3.23$\%$ over 12 test cases of 500 simulation steps. Figure \ref{fig:FL_32_fig1}(C) shows an example of physically accurate predictions made by the hybrid NN-PDE model. We omit the evaluation of baselines for high resolution cases as they do not succeed to model the flame dynamics for low resolution cases. 

In the following subsections, we present additional visualizations and a detailed comparison of the hybrid NN-PDE model against baseline approaches for different scenarios of Bunsen-type flames.

\subsubsection{uniform-Bunsen}
\begin{figure}
\centering
\includegraphics[width=0.9\linewidth]{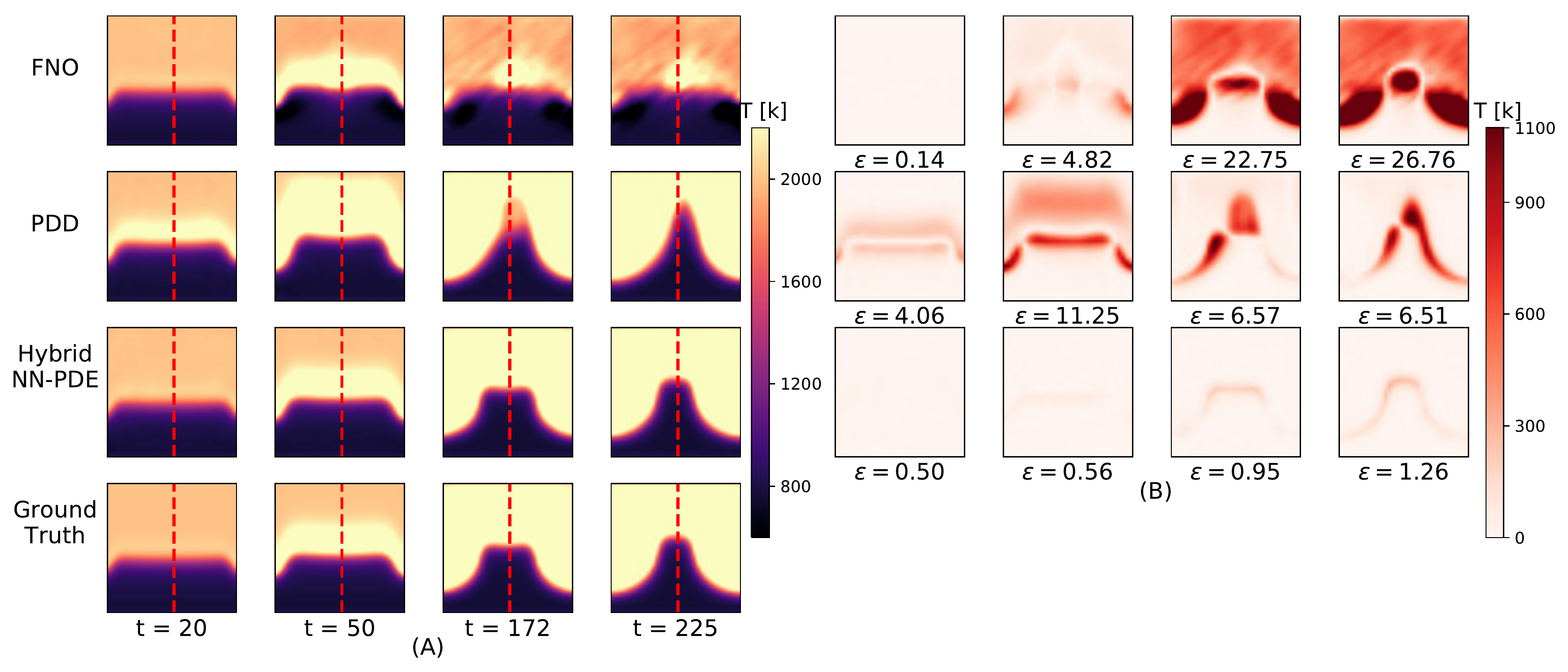}
\caption{uniform-Bunsen case with constant inlet velocity $U$ = 0.375 and $E$ = 0.95. (A) Temperature field predictions; (B) Absolute error between ground 
 truth ($\mathcal{P}_{c}(T_{t})$) and the output predicted - from top to bottom top - by: FNO, purely data-driven approach and hybrid NN-PDE. Hybrid NN-PDE model predicts physically accurate evolution of the flame cases under study}
\label{fig:fl_combined}
\end{figure}
Figures \ref{fig:fl_combined} show an example of the uniform-Bunsen case: a test case with $u_y = 0.375$ and $E = 0.95$. The constant inlet velocity results in a symmetric, $\Lambda$-shaped flame. Vertical, dotted, red line in figure \ref{fig:fl_combined}(A) helps to assess the symmetric nature of the flame. The purely data-driven model predicts a thicker flame ($t = 50$) or a flame with a spurious tip ($t = 172$) or an asymmetric flame ($t = 225$). Snapshots of the hybrid NN-PDE model predictions in figure \ref{fig:fl_combined}(A) show that it adapts to this scenario very well and succeeds in obtaining the correct results for long term forecasts of the temperature field. Furthermore, the flame shape and height are also better predicted, as shown in figure \ref{fig:fl_combined}(A). Very low error levels in figure \ref{fig:fl_combined}(B), such as $\epsilon = 1.26$ at $t = 300$, indicate that the hybrid model recovers the temperature field well.

\subsubsection{nonUniform-Bunsen}\label{subsec:nonuniform}
\begin{figure}
\centering
\includegraphics[width=0.9\linewidth]{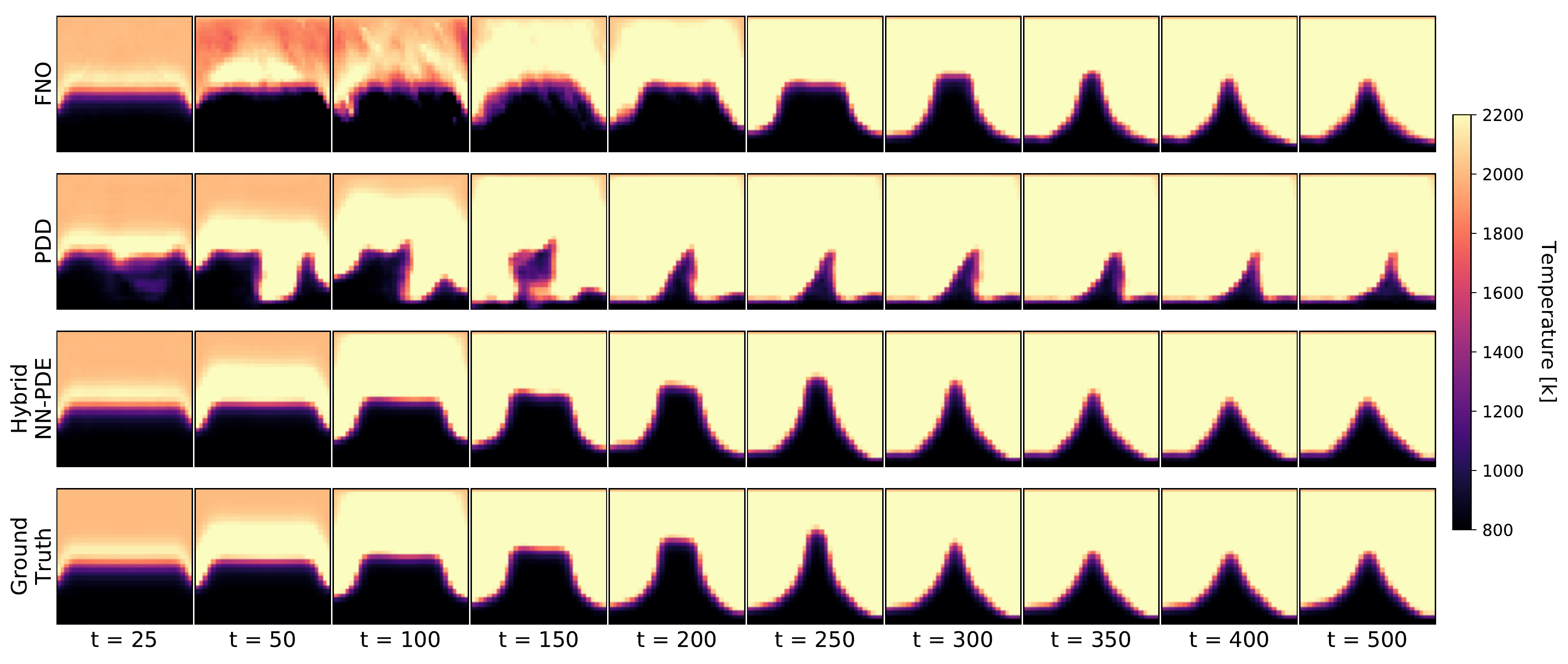}
\caption{nonUniform-Bunsen32 : Comparison between different approaches - from top to bottom - FNO, PDD, hybrid NN-PDE for a test case. Hybrid approach predictions accurately match with the ground truth data over long-rollouts of 500 time steps}
\label{fig:fl_nu_all}
\end{figure}
\begin{figure}
\centering
\includegraphics[width=0.9\linewidth]{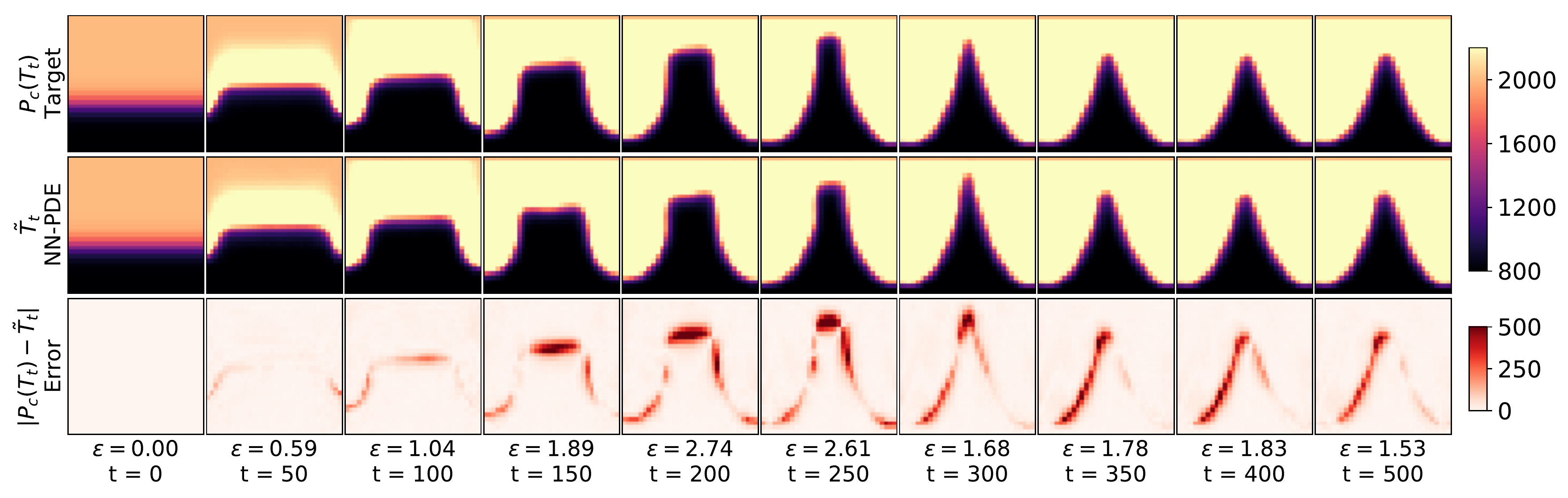}
\caption{nonUniform-Bunsen32 : Visualization of reactive flow trajectories predicted by the hybrid approach for different test case}
\label{fig:fl_nu_hyb}
\end{figure}
\begin{figure}
\centering
\includegraphics[width=0.9\linewidth]{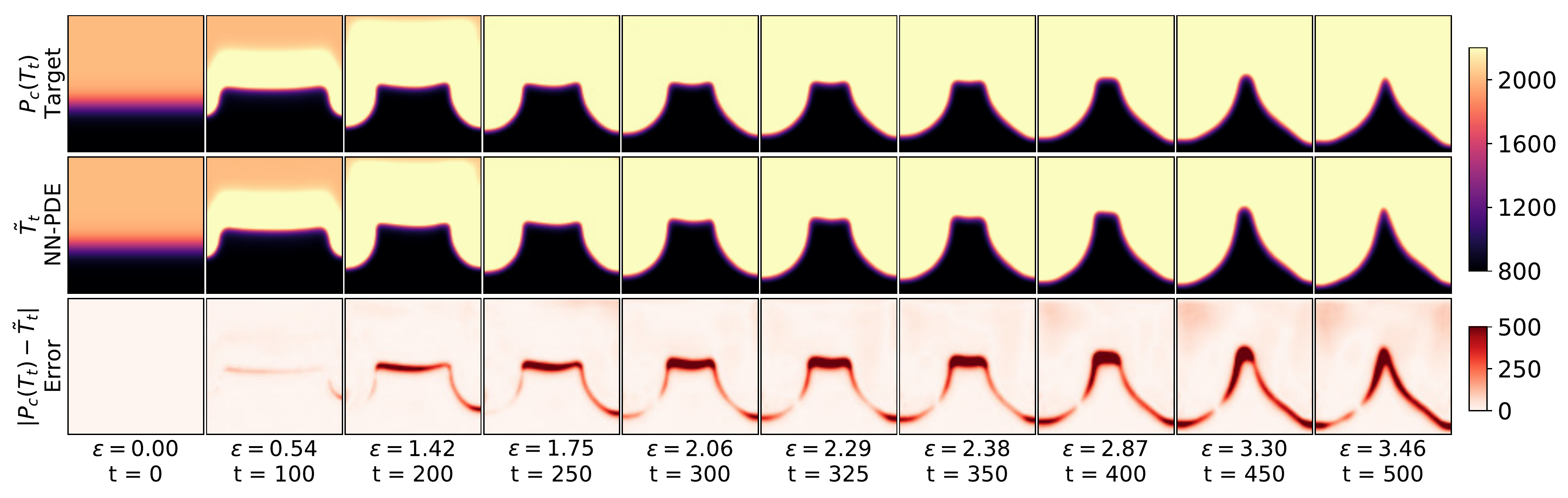}
\caption{nonUniform-Bunsen100 : Comparison between ground truth data ($\mathcal{P}_c(T_t)$) and hybrid NN-PDE approach predictions ($\tilde{T}_t$) for a different test case of complex, high resolution scenario}
\label{fig:fl_nu100}
\end{figure}

Figure \ref{fig:fl_nu_all} compares the temporal predictions made by FNO, PDD and the hybrid NN-PDE model with ground truth data for nonUniform-Bunsen32 case. Compared to Planar-v0 and uniform-Bunsen case, FNO predicts qualitatively better results for this case. Although still highly inaccurate, it predicts shapes that come closer to the target flame shapes for the later part of the simulation (for $t > 300$). This improvement might be due to the larger training dataset used for this scenario compared to the previous two scenarios. However, it fails to predict accurate temporal predictions over the entire simulation. The hybrid NN-PDE approach predicts the accurate flame evolution well. Additionally, we showcase the performance of the hybrid NN-PDE approach on two different test cases with complex flame shapes in figure \ref{fig:fl_nu_hyb}. The neural network model along with incomplete PDE solver reconstructs these complex flame shapes in an accurate manner.

For the nonUniform-Bunsen100 case, figure \ref{fig:fl_nu100} compares the predictions made by hybrid NN-PDE over $\{0, 100, 200, 250, 300, 325, 350, 400, 450, 500\}$ simulation steps with the ground truth data. It also shows the absolute difference between them. As the simulation progresses, higher errors are observed around the flame front. However, the hybrid NN-PDE approach captures the flame shape very accurately for longer roll-outs of 500 simulation steps. Currently this approach uses training dataset similar to nonUniform-Bunsen32. Further improvements in accuracy can be achieved by increasing the training dataset size or training the hybrid model with longer look-ahead steps. 

\section{Discussion}\label{sec:discussion}
In this section, we study the effect of temporal coarsening, incorrect PDE parameters and longer look-ahead steps on predictions made by the hybrid NN-PDE solver.

\subsection{Relaxing temporal stiffness in the PDE solver}\label{sec:vdt}
\begin{figure}
\centering
\includegraphics[width=0.75\linewidth]{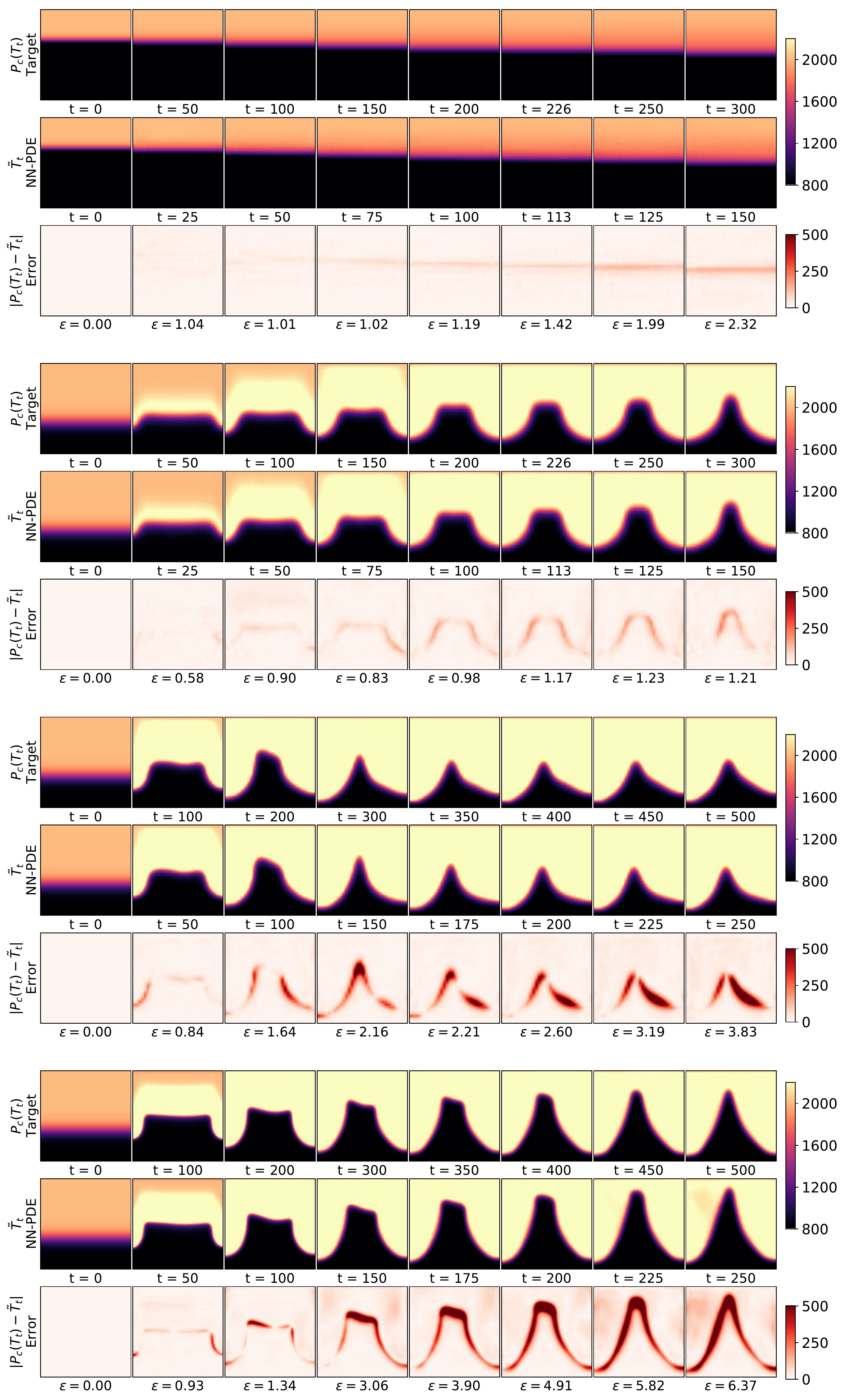}
\caption{Predictions made by hybrid NN-PDE model ($\tilde{T}_t$) with an incomplete PDE solver at twice the time-step of 2$\Delta t_c$, are compared with the ground truth solutions coming from the complete PDE solver at time-step of $\Delta t_c$. We showcase the effectiveness of hybrid approach in relaxing temporal stiffness of the complete PDE solver on reactive flow cases of - from top to bottom - Planar-v0, uniform-Bunsen, nonUniform-Bunsen32 and nonUniform-Bunsen100 for different test cases}
\label{fig:increased_dt_app}
\end{figure}
Traditionally, the source terms and reaction rate terms involved in modeling the fast chemistry of reacting flows require the use of very small time-steps in simulations due to the stiffness of the chemical mechanisms.
The incomplete PDE solver used in the hybrid NN-PDE approach, does not contain the source terms and reaction rate terms. Therefore, the time scales associated with the chemical reactions play a less important role in maintaining numerical stability, and it becomes possible for the solver to employ larger time steps. To illustrate this advantage, we train the neural network model with a time-step that is twice as large as the largest time-step $\Delta t_c$ required for the complete PDE solver to yield a numerically stable result. 

We mimic the setup described in section \ref{sec:data}, 
but now the incomplete PDE solver uses a time-step of $2 \Delta t_c$, which is too large for the complete PDE solver by itself to converge to a solution.
Figure \ref{fig:increased_dt_app} shows the results obtained from the models trained with a larger time-step are in good agreement with the target data for 4 different scenarios considered.
The flame dynamics are predicted accurately while using less simulation steps. 
For the uniform-Bunsen case, the hybrid approach takes $8.23 \pm 0.011 s$ to infer one simulation run, whereas the complete PDE solver requires $15.21 \pm 0.003 s$. Similar performance gains are observed across the other cases studied. 
Note that for the previous cases with a timestep size of one $\Delta t_c$, the trained model incurs a negligible runtime overhead. Even though incomplete PDE solvers are cheaper to run, as mentioned in the introduction section, increasing the spatial or temporal resolution of the incomplete PDE solver would not converge to the solutions of the complete description.

The last column of table \ref{tab:mape_mse} (\emph{Hybrid NN-PDE-dt}) summarizes the errors of the large time-step approach for all four reactive cases considered. Despite an effectively doubled computational performance, this model achieves similar errors to the hybrid NN-PDE approach. This highlights the capabilities of learned, hybrid PDE solvers, which can produce these solutions without the stability problems exhibited by the complete PDE solver, while at the same time being more accurate than pure data-driven predictions.

\subsection{Robustness of the hybrid solver}\label{sec:ablation}
\begin{figure}[bt!]
\centering
\includegraphics[width=0.9\linewidth]{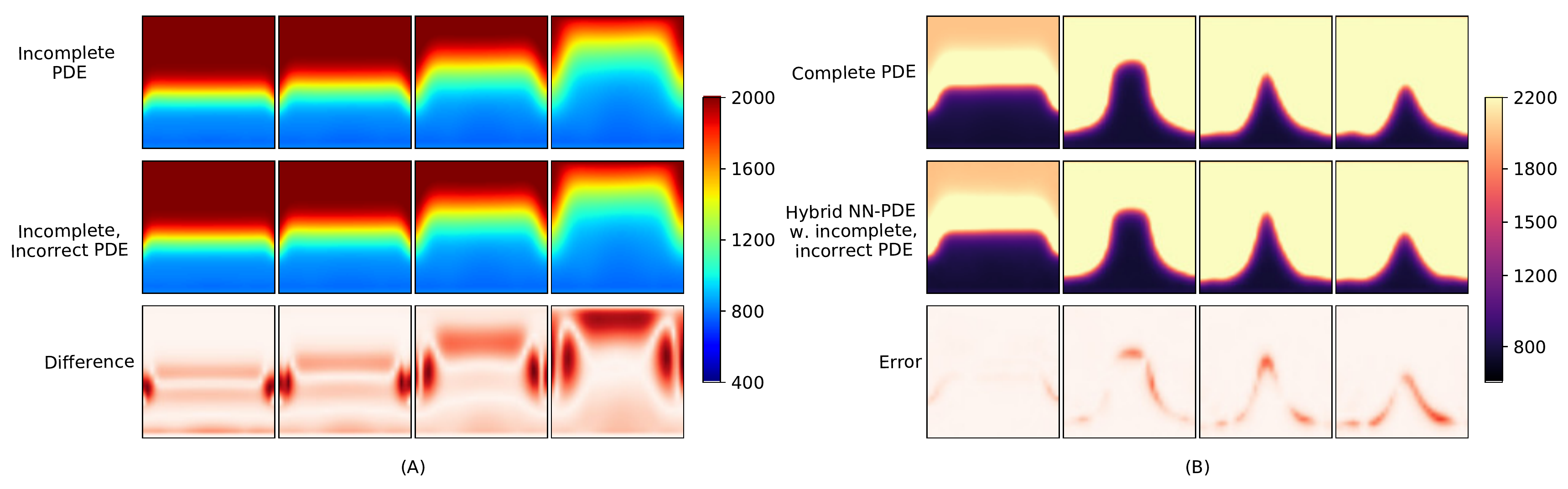}
\caption{Generalization to incorrect PDE parameters (A) visualization of differences in temperature fields due to incorrect parameters in incomplete PDE description. (B) Modified hybrid NN-PDE model, which combines the incomplete, incorrect PDE solver with neural network model is able to recover solutions of complete, correct PDE}
\label{fig:wrong_coeff}
\end{figure} 
\textbf{Generalization to incorrect PDE parameters \quad} In real-world scenarios, even if the incomplete PDE system is observable, one may have only an approximate understanding of the physics behind it. This may lead to incorrect estimation of the parameters in the incomplete PDE solver. The proposed hybrid NN-PDE model is capable of completing the PDE description even if the underlying incomplete PDE solver has incorrect parameters. We refer to the incomplete solver with incorrect parameters as ‘incomplete, incorrect PDE’. We experiment with a modified hybrid NN-PDE approach wherein we combine an incomplete, incorrect PDE solver with a neural network model. We assume that the known values of the incomplete PDE parameters in equation \eqref{ns_equations}, strain rate tensor ($\tau$), the diffusion coefficient of species $k$ ($D_k$) and mixture thermal conductivity ($\lambda$), are incorrect. Figure \ref{fig:wrong_coeff} (A) shows the difference between the temperature field evolution of the incomplete PDE solver with correct parameters ($[\tau, D_k, \lambda] = [0.1, 0.1, 0.1]$) and incorrect parameters ($[\tau, D_k, \lambda] = [0.05, 0.05, 0.05]$) at different time-steps. The hybrid NN-PDE combines this incomplete, incorrect PDE solver with the neural network model to obtain the solutions of the complete PDE solver with correct parameters. Figure \ref{fig:wrong_coeff} (B) compares the flame dynamics predicted by this hybrid NN-PDE model with that of the complete, correct PDE solver. A good match is observed over various test cases. The hybrid model with incomplete, incorrect solver achieves an overall MAPE of 2.48 $\pm$  1.20 \%, compared to the MAPE of 2.04  $\pm$ 1.39 \% for hybrid model with incomplete, correct PDE.

\begin{figure}[bt!]
\centering
\includegraphics[width=0.9\linewidth]{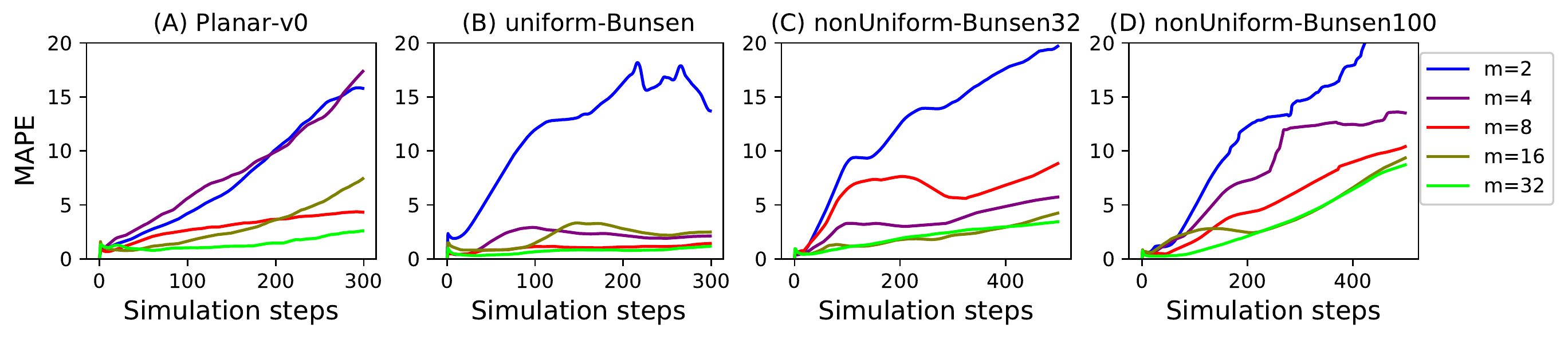}
\caption{Effect of longer look-ahead steps. MAPE of temperature field predictions by hybrid NN-PDE model, over all testsets. Models trained with higher look-ahead steps $m$ accurately predict the temporal evolution of dynamics for longer duration across all cases considered.}
\label{fig:look_ahead}
\end{figure}
\textbf{Effect of longer look-ahead steps \quad} We evaluate the effect of varying look-ahead steps on the performance of the hybrid model. We show the comparison of models trained with $m = \{2,4,8,16,32\}$ for different cases in figure \ref{fig:look_ahead}. Models with smaller $m$ (2) do not learn to accurately correct the fields over long time and quickly diverge from the target simulation. Using larger $m$ improves the quality of prediction drastically as the model learns the correction via the gradients over longer simulation steps. For Planar-v0 and uniform-Bunsen cases, iterating the NN and PDE solver for 32 time-steps improves the accuracy by 81.7\% and 94\% respectively compared to the $m = 2$ model. For nonUniform-Bunsen32 and nonUniform-Bunsen100, the model performance improves by 84.2\% and 70.9\% respectively, by using $m = 32$ instead of $m = 2$ model.

For some of error curves in figure \ref{fig:look_ahead}, significant error fluctuations are observed for lower simulation steps followed by recovery as the simulation progresses (e.g. figure \ref{fig:look_ahead} (C), $m=4$). The evolution of the flame dynamics over time is highly nonlinear owing to the strong nonlinearity of the system where the premixed fuel and air mixture reacts to form the products. During the initial steps of this transient behaviour, the distribution of flame temperature and flame interface varies rapidly. This leads to steep temperature gradients across the computational domain. As the simulation approaches towards the steady-state flame shape, the temperatures across the domain are in a similar range across different datasets. Therefore, small errors in flame temperature predictions and flame velocity predictions may lead to larger errors during the transient phase of the flame before it reaches a steady-state.

\begin{figure}[bt!]
\centering
\includegraphics[width=0.75\linewidth]{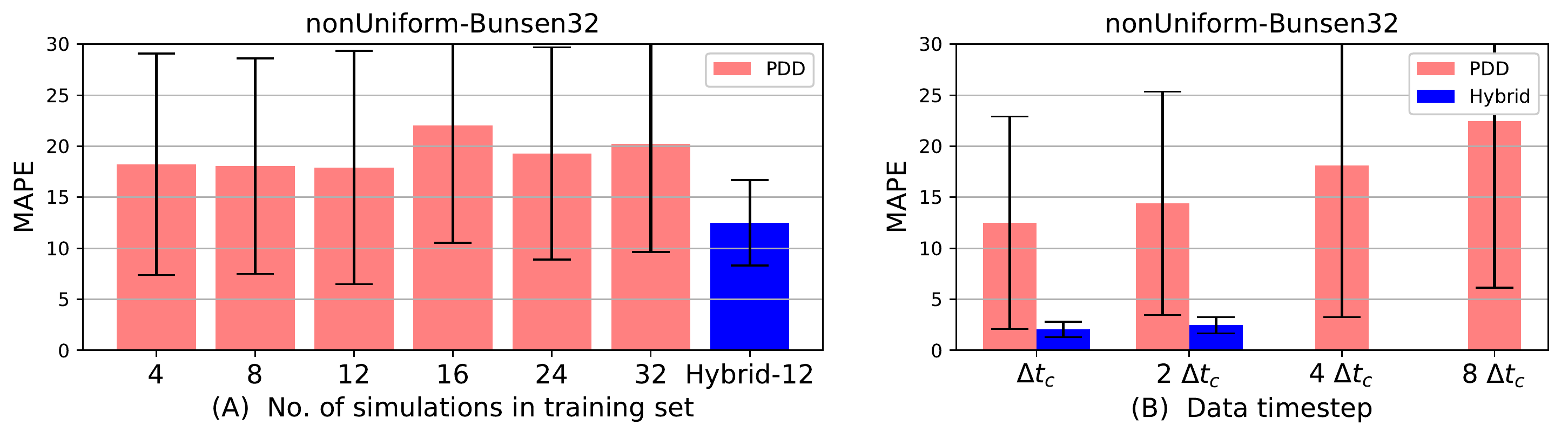}
\caption{(A) Effect of increasing training dataset size for the PDD approach, over fixed testsets, is compared with equivalent (trained with same look-ahead steps $m = 2$) hybrid NN-PDE model. (B) Effect of temporal coarsening on PDD models trained with $m = 32$ look-ahead steps}
\label{fig:histogram}
\end{figure}

\textbf{Effect of training dataset size \quad} We study the effect of training dataset size on the accuracy of PDD predictions. We train PDD models with different numbers of simulations in the training dataset. Each simulation contains 500 time-steps. Figure \ref{fig:histogram} (A) shows the MAPE of PDD models trained with $\{4, 8, 12, 16, 24, 32\}$ training datasets, over a fixed testset. We compare the performance of these PDD models with an equivalent (trained with same look-ahead steps $m=2$) hybrid NN-PDE model, trained with 12 simulation sets. Figure \ref{fig:histogram} (A) shows increasing the number of training sets has little or no effect on the prediction capabilities of the PDD models for nonUniform-Bunsen32 case. The hybrid model with 12 training sets achieves a MAPE of 12.49 $\pm$ 4.17\%, an improvement of 38\% over the PDD model with 32 training sets. The purely data-driven models cannot achieve the same level of accuracy even in the presence of large amounts of data. This result strengthens the hypothesis that integrating the incomplete PDE solver into the neural network training yields a learning signal that fundamentally differs from that produced by training with precomputed data. The purely data-driven models cannot achieve the same level of accuracy even in the presence of large amounts of data.
\begin{figure}[bt!]
\centering
\includegraphics[width=0.8\linewidth]{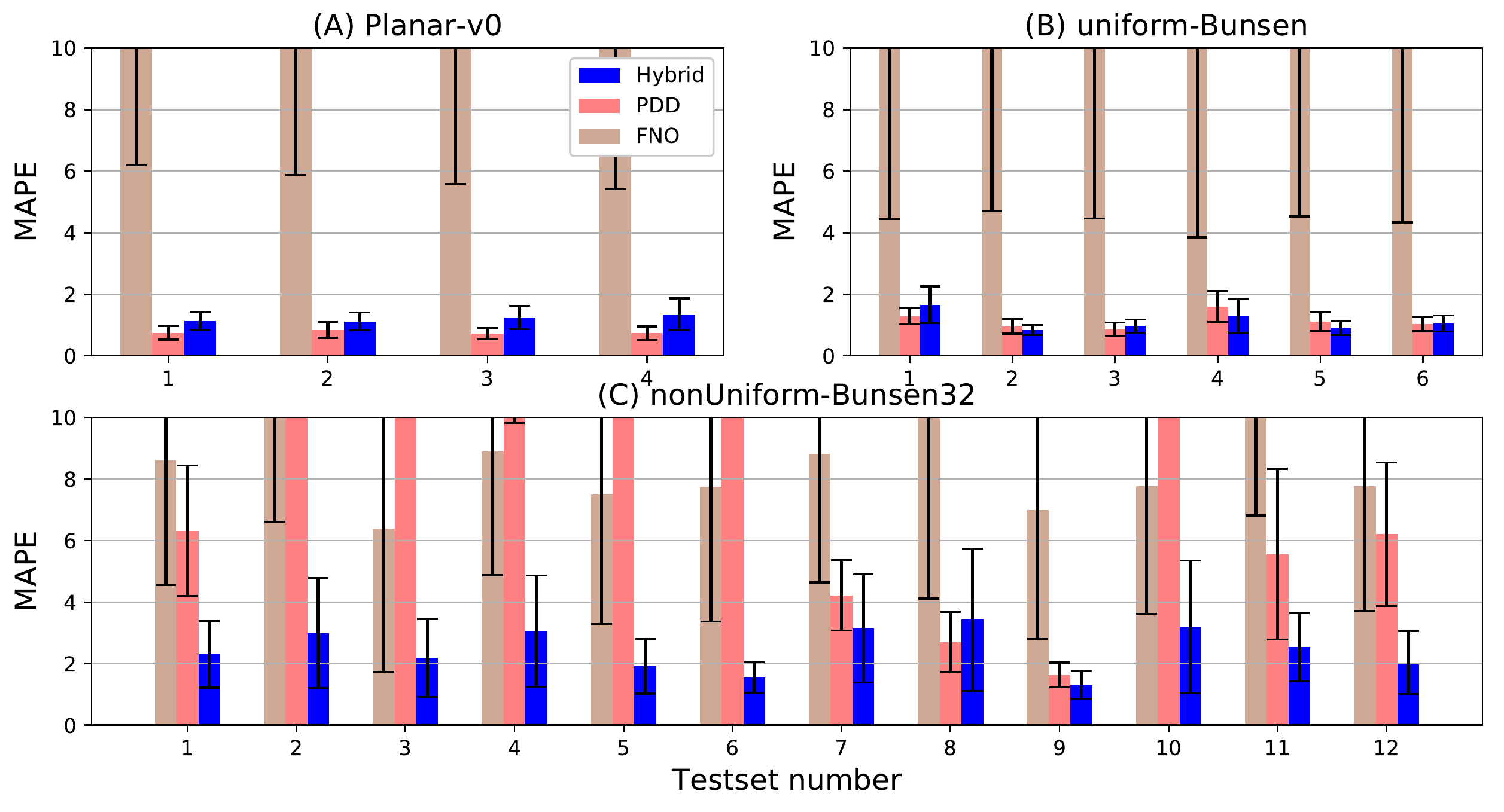}
\caption{Bar plot of MAPE of temperature field predictions by FNO, PDD and hybrid NN-PDE model at time-step $2 \Delta t_c$, for different testcases}
\label{fig:hist_mape_vdt}
\end{figure} 

\begin{figure}
\centering
\includegraphics[width=0.8\linewidth]{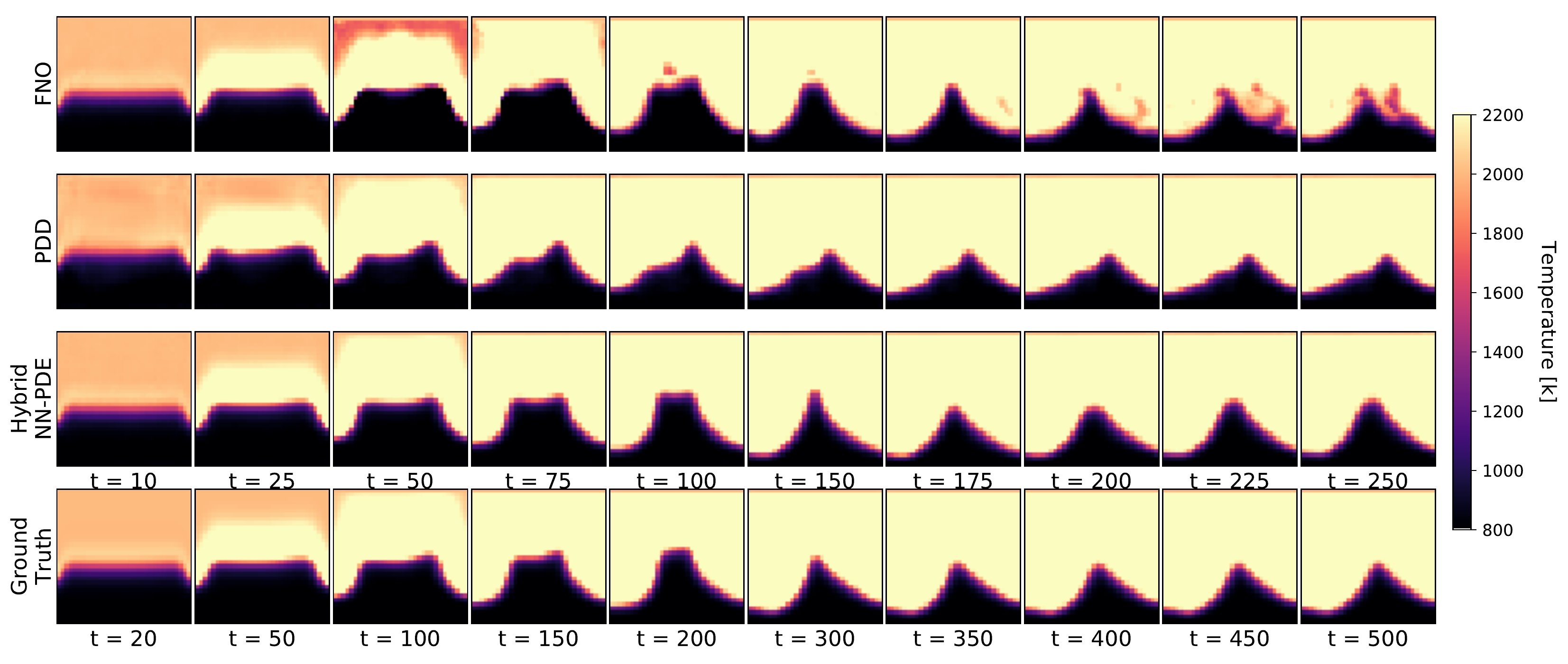}
\caption{Comparison between different approaches at time-step $2 \Delta t_c$ - from top to bottom - FNO, PDD, hybrid NN-PDE for nonUniform-Bunsen32 test case. Hybrid approach predictions accurately match with the ground truth data over long roll-outs}
\label{fig:fl_nu_dt_all}
\end{figure}
\textbf{Effect of temporal coarsening \quad}
Table \ref{tab:mape_mse} compare the MAPE of temperature field predictions at time-step $\Delta t_c$, for all three scenarios. The hybrid NN-PDE approach consistently performs better than both the baselines for all three scenarios considered. Higher standard deviations in table \ref{tab:mape_mse} indicate the large differences in the prediction of flame temperatures for baseline approaches. The multi-physics systems we study provide a substantially more difficult environment than regular fluids: the chemical reactions are numerically very stiff, and the resulting dynamic interactions are difficult to capture by numerical solvers. To further illustrate the temporal behavior of the simulations, we perform additional experiments with the baseline approaches using twice the time-step size of the complete solver ($2 \Delta t_c$). As shown in figure \ref{fig:hist_mape_vdt}, for the Planar-v0 and uniform-Bunsen case, PDD catches up with the hybrid approach whereas the PDD approach completely fails to predict the dynamics of the nonUniform-Bunsen32 case. 
We further investigate the performance of the PDD approach on the nonUniform-Bunsen32 case with larger time-steps (2 times, 4 times and 8 times) as shown in the figure \ref{fig:histogram} (B). PDD achieves higher MAPE of 22.44 \% for 8$\Delta t_c$ setup as compared to the MAPE of 12.48 \% for $\Delta t_c$ setup. The PDD approach does not predict the dynamics accurately for any of the larger time-steps considered. We can not showcase the performance of the hybrid NN-PDE solver for time steps greater than $2 \Delta t_c$ as it is restricted by the underlying incomplete PDE solver.
Figure \ref{fig:fl_nu_dt_all} visualize these results over different time-steps for one of the test cases. The predictions made by FNO, PDD and hybrid NN-PDE model for nonUniform-Bunsen32 case at twice the time-step of $2 \Delta t_c$, are compared with the ground truth solutions coming from the complete PDE solver at time-step of $\Delta t_c$. FNO and PDD (upper two rows) fail to recover the correct flame shapes over 250 simulation steps. The hybrid NN-PDE model (second row from below) predicts the flame shape accurately, thus relaxing the temporal stiffness of the complete PDE solver.

\section{Flame shape control}\label{sec:flame_shape}
To demonstrate the flexibility of the hybrid NN-PDE solver of the previous sections we consider a multi-physics system control problem. We test the joint applicability of a neural network with a pre-trained hybrid reactive flow solver
to obtain a desired flame shape. We integrate the hybrid solver as a layer into the neural network model to enable the training of a flame shape controller. As a result, physics knowledge from the hybrid solver can be embedded into the optimization problem, by matching observations coming from the solver. The hybrid solver provides the agent with feedback of how interactions at any point in time affect the flame shape. We consider a physical system of reactive flows where the agent can interact with the system by controlling the inlet velocity of fuel-air flow.
\begin{figure}[bt!]
\centering
\includegraphics[width=0.8\linewidth]{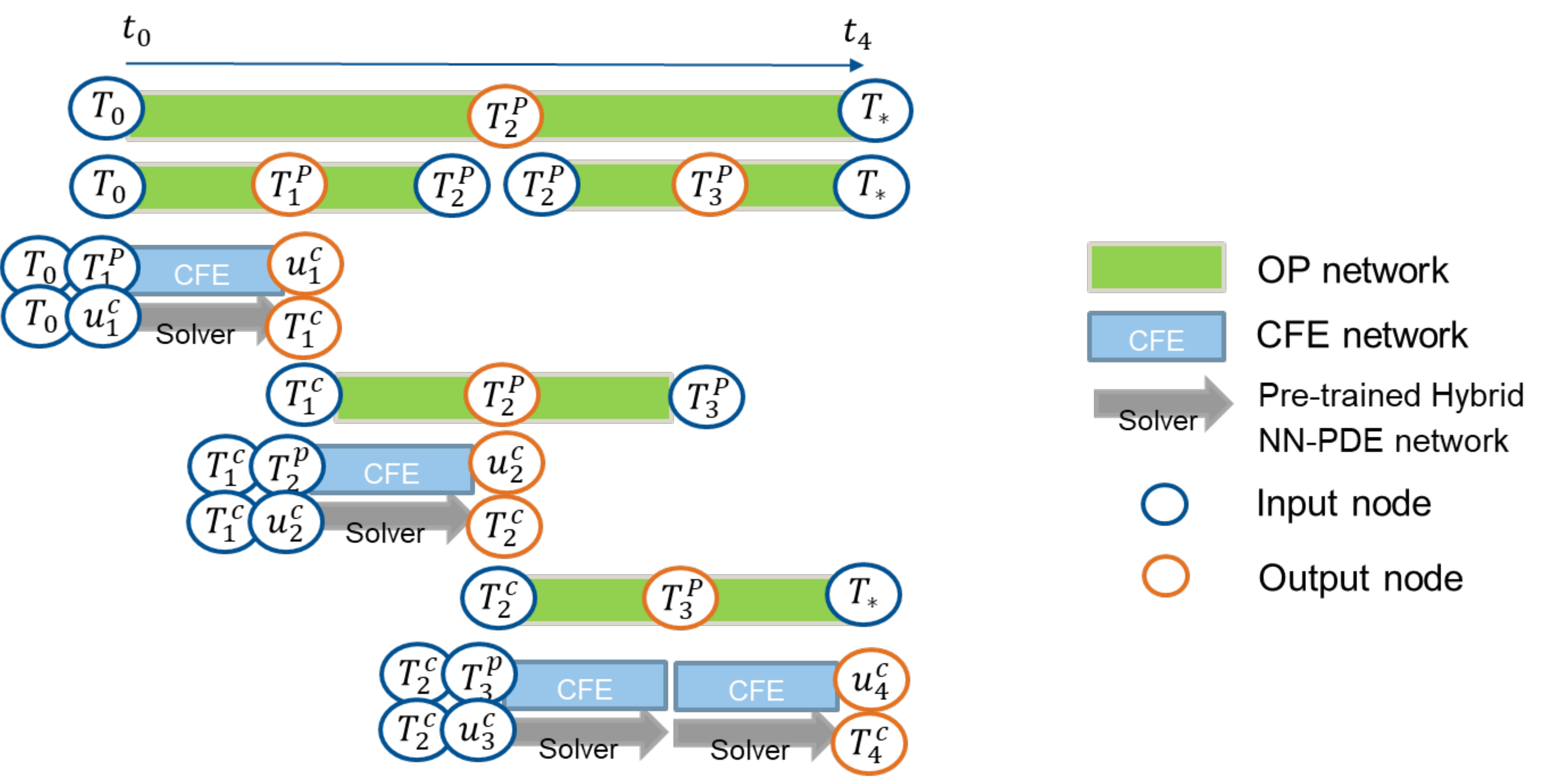}
\caption{Details of the OP-CFE prediction sequence}
\label{fig:op_cfe}
\end{figure}

We use a predictor-corrector approach \citep{holl2019pdecontrol} with a supervised loss function and control sequence refinement for the nonUniform-Bunsen case with 32 $\times$ 32 resolution. A pre-trained hybrid solver from section \ref{subsec:nonuniform} is used as a reactive flow solver. This hybrid model consists of incomplete PDE solver along with an UNet model detailed in section \ref{sec:models}, section \ref{sec:training_details} and figures \ref{fig:solvernn}, \ref{fig:res_unet}. It is first trained using training instances of the uniform-Bunsen32 case described in section \ref{sec:data}. Additionally, it is further trained on the training data of flame shape transitions from an arbitrary flame shape to any other arbitrary flame shape. Once trained, the weights of the hybrid solver are frozen and used as a layer into the predictor-corrector approach.
It is differentiable by construction, when combined with the deep neural networks is shown to learn a policy to control flame shapes. The learning objective is to arrive at a target temperature field ($T_*$) with the desired flame shape from an arbitrary, given initial configuration ($T_0$) by controlling the velocity field ($u_i^c$). We consider a real-world setting where only few states of the reactive flow system are observable and controllable: the temperature and mass-fraction fields are observable, and only the inlet of the velocity field is controllable. The predictor-corrector approach consists of two neural network models. First, an observation predictor network (OP), predicts the intermediate states of the observable quantity i.e. temperature in this case ($T_i^p$). The corrector (CFE) network predicts the control force ($u_i^c$) required to follow the trajectory predicted by OP network. The inlet velocity field predicted by the CFE network is used by the pre-trained hybrid solver to compute the corrected temperature field ($T_i^c$). Figure \ref{fig:op_cfe} shows the sequence of OP-CFE network used to control the flame shape $T_*$ given the initial flame shape $T_0$, over the trajectory of 4 time-steps.
The OP network is modelled as a temporal hierarchical process to incorporate the knowledge about longer time spans. Therefore, OP networks are trained to predict the optimal center point between two time steps instead of predicting the state of the next time step. The recursive call to the OP network enables the prediction of observable quantities at every time step ($T_i^p$). Using $T_{i-1}$ and OP prediction at step \textit{i}, i.e. $T_i^p$, CFE network predicts the control force at step i, $u_i^c$. Once $u_i^c$ is predicted, learned hybrid solver can be used to advance the simulation to next time step and obtain the actual state $T_i^c$. Using this sequence of OP-CFE network, we can obtain $T_n^c$ with $n$ CFE evaluations, which matches target flame shape $T_*$ very closely. This sequence can be generalized to any arbitrarily long length of sequences. OP-CFE networks are trained using the supervised loss function on the target temperature field achieved and inlet velocity predicted, respectively. The loss function formulations for OP and CFE network are,
\begin{equation}
    L_{op} = |OP[T_i, T_j] - T^{GT}(\frac{i+j}{j})|^2 \quad \quad L_{CFE} = \sum_{i=1}^n |u_i^c - u_i^*|^2,
\end{equation}
 where $T^{GT}$ and $u^*$ indicate the ground truth temperature field and inlet velocity, respectively. The simplest technique to obtain the optimal trajectory, given an initial state is to use the chain of CFE networks followed by the solver. CFE, modeled using deep neural network, predicts the control force required to predict the transition to next observable state. Therefore, as a baseline model, we consider the CFE only approach. 
 All neural networks used in this work are modified UNet architectures \citep{ronneberger2015u}.

\begin{figure}[bt!]
\centering
\includegraphics[width=0.8\linewidth]{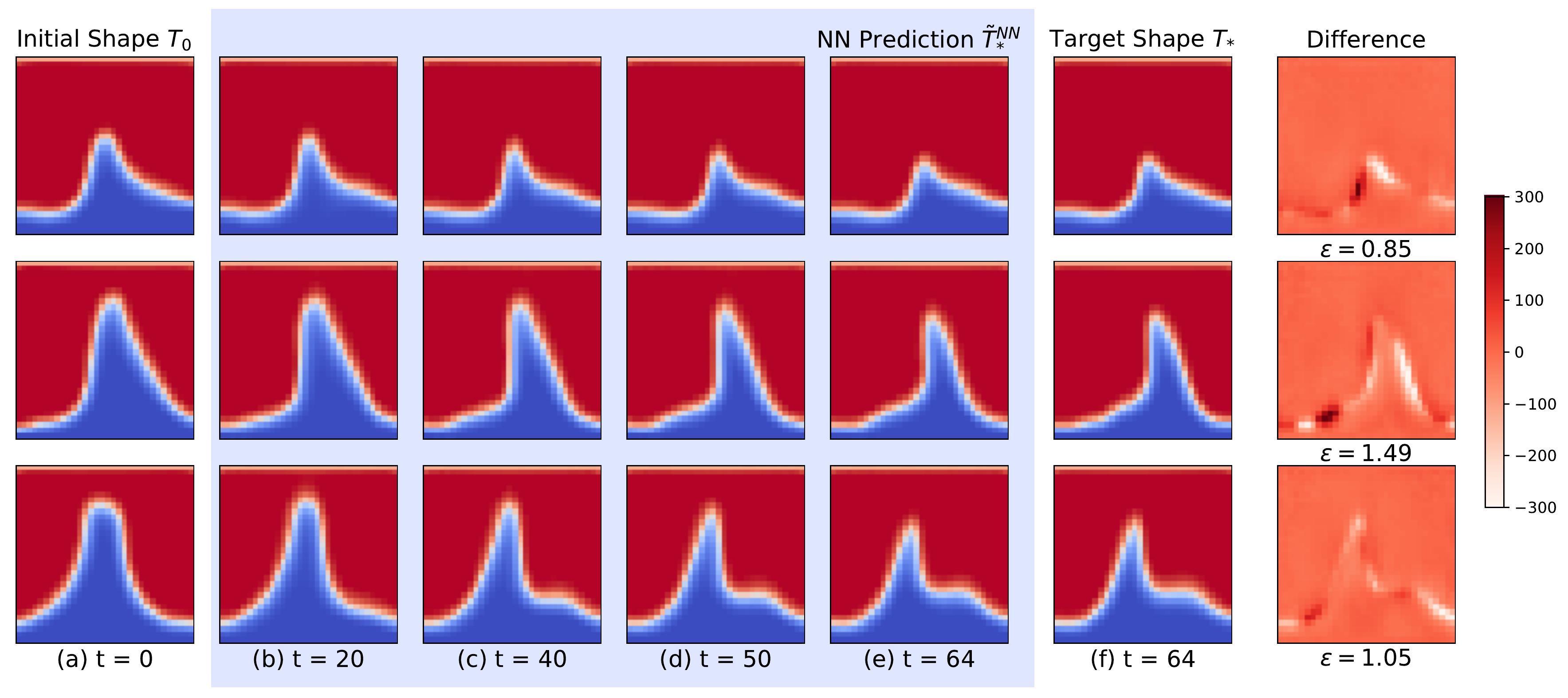}
\caption{Temperature field predictions achieved by the neural network model $\tilde{T}_{*}^{NN}$ from given initial flame shape $T_0$ to achieve target flame shape $\tilde{T}_{*}$ by controlling flow velocity}
\label{fig:control}
\end{figure}

Figure \ref{fig:control} (b-e) shows examples of the transition achieved and compares the target flame shape obtained using the neural network model with the target flame shape ($T_*$). All three cases are test cases, i.e. were not seen at training time, and have a sequence length of 64 simulation steps. 
It is visible that the neural network model succeeds at controlling the inlet velocity of fuel-air mixture flow to obtain the desired flame shapes. To quantify these results, the learned model achieves MAPE of 1.34 $\pm$ 0.41\% over 50 test cases considered. It achieves an improvement of 65.7\% compared to a baseline model of the CFE only approach, without a predictor network for the long term prediction of temperature fields. This baseline model has a MAPE of 3.91 $\pm$ 2.19\%. This case highlights the capabilities of differentiable hybrid NN-PDE multi-physics solvers. Specifically, the hybrid solver helps the OP-CFE networks to yield controllers that steer the complex physics of the reactive flow over long time spans.

\section{Conclusion}\label{sec:conclusion}
In this work, we proposed a hybrid NN-PDE solver which learns to complete an incomplete PDE solver in the context of a multi-physics system, namely the case of a reacting flow. The trained neural network model works alongside an incomplete solver to accurately account for and predict the effects of unknown physics. We compare its performance with two baselines - a purely data-driven approach and state-of-the-art FNO approach. 
We test it on the cases of reacting flows of increasing complexity.
The qualitative and quantitative performance of the presented hybrid NN-PDE approach is found to be superior compared to these baselines to predict the long-term temporal evolution of the reactive flows.
The incomplete PDE description helps the neural network model recover the target simulation with a significantly improved accuracy. This hybrid NN-PDE model is able to predict the correct evolution of the important features of the multi-physics system (in our case, the flame interface and flame shape) for longer simulation steps in all scenarios and is able to generalize to other (initial and boundary) conditions of the system. This is demonstrated by variations in the equivalence ratio and inlet velocity forcing. We also show that such hybrid approach may have the added benefits of allowing to relax numerical constraints linked to the potential stiffness of the unknown physics. The learned hybrid NN-PDE solver is successfully adapted to enable flame shape control by combining with deep neural networks.

Our work represents a stepping stone for numerous avenues of future work. While we demonstrated the applicability of the proposed approach to the complex case of a reacting flow (with  strong nonlinear interactions between chemistry and velocity), the proposed hybrid NN-PDE solver could be further utilized to predict and control the dynamics of other tightly coupled multi-physics systems \citep{levy1999two, dowell2001modeling}. In addition, applying the proposed approach to other cases, where a limited amount of real data is available would be a very interesting  step. 
Especially if the data is very sparse and contains noise, learning meaningful corrections could become excessively challenging. Performing such an assessment will be explored in future work.
We note that the proposed approach does not seek to replace the classical physical simulation approaches. However, it is an important step in the direction of harnessing the capabilities of neural network models using the knowledge of partial differential equations for complex multi-physical systems.

\section{Acknowledgements}
N. Tathawadekar acknowledges the financial support of the European Union’s Horizon 2020 research and innovation program under the Marie Sklodowska-Curie grant agreement No. 766264 and the ERC Consolidator Grant \textit{SpaTe} (CoG-2019-863850).

\section{Data Availability Statement}
The code to generate the data and train models that support the findings of this study are openly available at \url{https://github.com/tum-pbs/Hybrid-Solver-for-Reactive-Flows}.


\appendix
\setcounter{figure}{0}   
\setcounter{table}{0}  
\section{Appendix}
\subsection{Choice of NN architecture}\label{app:NN}
\begin{table}[b!]
\caption{Details of various neural network architectures used in this paper.}
\label{tab:nn_details}
\small
\begin{center}
\begin{tabular}{ |c|c c c| } 
\hline
\\[-1em]
Hyper-parameters &  ResNet & UNet32 & UNet100 \\
\hline
Kernel size & 5 & 5 & 5 \\
Latent size & 32 & & \\
Activation & LeakyReLu & LeakyReLu & ReLu \\
Loss & MSE & MSE & MSE \\
\# ResBlocks & 5 & & \\
\# UNet layers & & 2 & 3 \\
CNN stack depth & 2 & 3 & 2 \\
Base latent size & & 16 & 16 \\
Spatial down-sample by layer &  & 2 & 2 \\
latent sizes & & (32) & (32,64) \\
\# trainable parameters & 261,953 & 136,227 & 362,371 \\
\hline
\end{tabular}
\end{center}
\end{table}

\begin{table}[bt!]
\caption{Mean and standard deviation of errors over different architectures for purely data-driven and hybrid NN-PDE approaches.}
\label{tab:unet_resnet}
\begin{center}
\resizebox{0.95\textwidth}{!}{
\begin{tabular}{ c|c|c|c||c|c| } 
\cline{2-6}
\\[-1em]
& & PDD & PDD & Hybrid NN-PDE & Hybrid NN-PDE \\
& & ResNet & UNet & ResNet & UNet \\
\cline{2-6}
\\[-1em]
\parbox[t]{2mm}{\multirow{4}{*}{\rotatebox[origin=c]{90}{MAPE}}} & Planar-v0 & 6.33 $\pm$ 3.05\% & 7.09 $\pm$ 4.40\% & 1.62 $\pm$ 0.44\% & \textbf{1.40} $\pm$ 0.65\% \\
& uniform-Bunsen & 7.58 $\pm$ 3.73\% & 2.87 $\pm$ 1.53\% & 2.01 $\pm$ 0.99\% & \textbf{0.72} $\pm$ 0.37\% \\
& nonUniform-Bunsen32 & 12.48 $\pm$ 11.31\% & 19.19 $\pm$ 14.92\% & 3.25 $\pm$ 2.35\% & \textbf{2.04} $\pm$ 1.39\%  \\
& nonUniform-Bunsen100 & - & - & 15.68 $\pm$ 10.44\% & \textbf{3.23} $\pm$ 3.76\%  \\
\cline{2-6}
\end{tabular}}
\end{center}
\end{table}
We experimented with ResNet and UNet architectures for the neural network part of the purely data driven and hybrid NN-PDE approach discussed in section 3.3 of the main paper. 
Table \ref{tab:nn_details} provides details of the convolutional blocks, layers of the network, and activation function. The \emph{UNet32} architecture with 2 layers is used for 32 $\times$ 32 cases and the \emph{UNet100} architecture with 3 layers is used for 100 $\times$ 100 case discussed in the paper. Table \ref{tab:unet_resnet} shows the comparison between MAPE achieved by ResNet and UNet architectures for the purely data-driven and the hybrid NN-PDE approach for different cases considered in the paper. For the hybrid NN-PDE approach with 32 $\times$ 32 resolution cases, the ResNet and UNet achieve comparable performance with the UNet yielding the lowest error. The main difference arises in the 100 $\times$ 100 resolution case. The ResNet fails to converge during training, leading to high test error. On the contrary, the UNet achieves low training and testing errors. The main advantage of UNet comes from transforming high resolution input to low resolution representation and reconstructing it back to the high resolution output. Therefore, we choose the UNet architecture for the hybrid NN-PDE approach. 
For the purely data-driven approach the ResNet was found to be better for 2 out of 3 cases. Specifically for the nonUniform-Bunsen32 case, UNet performs very poorly for the PDD approach. Therefore, we report ResNet errors for the PDD approach throughout the paper. 

\subsection{Additional Experiments}\label{app:PDD_5}
\begin{figure}
\centering
\includegraphics[width=0.9\linewidth]{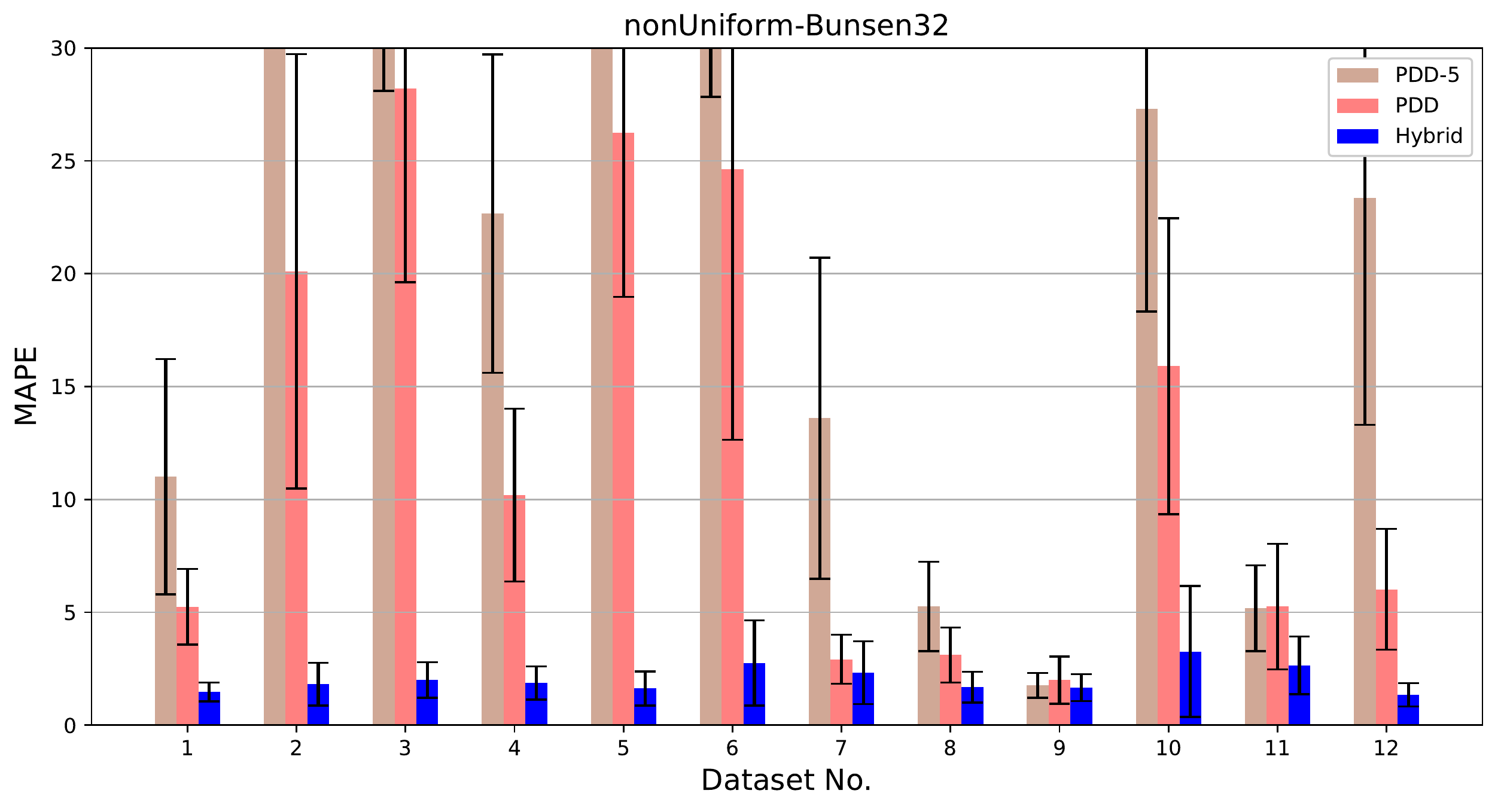}
\caption{Bar plot of MAPE of temperature field predictions by a purely data-driven model trained to predict 5 quantities of flow state (PDD-5), PDD model trained to predict 3 quantities of flow state (PDD), and hybrid NN-PDE model (Hybrid)}
\label{fig:mape_ppd_5r}
\end{figure}

We additionally compare against the variant of the purely data-driven model described in section 3.2. It employs a neural network model to learn the complete flow states $\mathcal{P}_c(\phi)$ given an input flow state $\phi^S$ where $\phi^S = [T, Y_f, Y_o, u, p]$. Keeping all hyperparameters the same, the ResNet is trained to predict all quantities of flow state for nonUniform-Bunsen32 case. 
As the PDD approach does not enforce any physical laws, the inclusion of additional quantities results in a more difficult learning task, which given the same number of learnable parameters naturally results in a deteriorated quality of the inferred results.
Figure \ref{fig:mape_ppd_5r} shows the MAPE of the temperature field predicted by this new variant (referred to as PDD-5) with the PDD approach described in section 3.3 and hybrid NN-PDE approach for the nonUniform-Bunsen32 case. Across all test cases, PDD-5 performs worse with an overall MAPE of 23.45 $\pm$ 18.07\% than the PDD approach (12.48 $\pm$ 11.31\%) as problem complexity has increased. 


\bibliography{jfm.bib}

\end{document}